\documentclass[useAMS,usenatbib]{mnras}
\usepackage{graphicx}
\usepackage{amsmath}
\usepackage{amssymb}

\def\gsim{\;\rlap{\lower 2.5pt
 \hbox{$\sim$}}\raise 1.5pt\hbox{$>$}\;}
\def\lsim{\;\rlap{\lower 2.5pt
   \hbox{$\sim$}}\raise 1.5pt\hbox{$<$}\;}
   
\newcommand{\tr}[1]{\textrm{#1}}
\newcommand{\ee}[1]{\times10^{#1}}
\newcommand{\dd}[2]{\frac{\textrm{d}#1}{\textrm{d}#2}}
\newcommand{\pp}[2]{\frac{\partial#1}{\partial#2}}

\newcommand{\mb}{\mathbf}

\title{Constraints on Cosmic Ray Transport in Galaxy Clusters from Radio and Gamma Ray Observations}
\author[Wiener \& Zweibel]{Joshua Wiener$^{1}$ \& Ellen G. Zweibel$^{1,2}$\\
$^{1}$ Department of Astronomy, University of Wisconsin-Madison, Madison, WI 53706, USA.\\
$^{2}$ Department of Physics, University of Wisconsin-Madison, Madison, WI 53706, USA.}

\begin{document}
\bibliographystyle{mnras}

\maketitle

\label{firstpage}

\begin{abstract}
The nature of cosmic rays (CRs) and their transport in galaxy clusters is probed by several observations. Radio observations reveal synchrotron radiation of cosmic ray electrons (CRe) spiraling around cluster magnetic fields. $\gamma$-ray observations reveal hadronic reactions of cosmic ray protons (CRp) with gas nuclei which produce pions. No such cluster-wide $\gamma$-ray signal has been measured, putting an upper limit on the density of CRp in clusters. But the presence of CRe implies some source of CRp, and consequently there must be some CRp loss mechanism. We quantify the observational constraints on this mechanism assuming that losses are dominated by CR transport, ultimately deriving lower limits on this transport. Using the Coma cluster as an example, we find that bulk outward speeds of 10-100 km s$^{-1}$ are sufficient to reduce $\gamma$-radiation below current upper limits. These speeds are sub-Alfv\'enic and are consistent with a self-confinement model for CR transport if the magnetic field is coherent on large scales. If transport is diffusive, we require minimum diffusion coefficients of 10$^{31}$-10$^{32}$ cm$^2$s$^{-1}$. This is consistent with CRs free streaming at the speed of light along a field tangled on length scales of a few kpc. We find that a model of the Coma cluster with a tangled field and the self-confinement picture together can be consistent with observations if the relative acceleration efficiency of CR protons is less than 15 times more than electrons of the same energy. This value is 3-6 times lower than the same quantity for Galactic cosmic rays.
\end{abstract}

\section{Introduction}
Galaxy clusters are the largest ($\sim 10^{14}-10^{15}M_\odot$) gravitationally bound objects in the Universe and are host to a plethora of physical processes ranging orders of magnitude in scale length and energy. The energy budget of clusters includes several non-thermal components, such as magnetic fields and cosmic rays (CRs), which are potentially important in cluster dynamics. For instance, wave heating by CRs has been proposed as a heating mechanism to prevent cooling catastrophes in cool core (CC) clusters (\cite{loewenstein91,Guo08a}).

Some information about the magnetic and CR content of clusters can be determined from radio observations, which in some clusters reveal large scale diffuse synchrotron emission from cosmic ray electrons (CRe). These giant radio halos provide insight into the nature of cluster-wide magnetic fields and CR transport. The Coma radio halo in particular has been the target of several radio observations and studies (see \cite{deiss97}, \cite{thierbach03}, \cite{brown11-coma}, and \cite{brunetti12} for just a few examples).

Cosmic ray protons (CRp) can be independently detected from their hadronic reactions - neutral pions produced in high energy hadronic collisions decay into $\gamma$-rays (charged pions also produced in these collisions decay into CR electrons and positrons, referred to as `CR secondaries'). Since $\gamma$-radiation can also be produced as inverse Compton emission from high energy electrons up-scattering cosmic microwave background (CMB) photons, any detection of $\gamma$-rays is only an upper bound on the rate of hadronic reactions. However, while $\gamma$-radiation has been seen in several galaxies, no diffuse cluster-wide $\gamma$-ray emission has yet been detected despite deep searches with $\gamma$-ray telescopes such as the \emph{Fermi} Large Area Telescope (see \cite{huber13} and \cite{fermi16} for just two examples). These non-detections put strict upper limits on the CRp content of galaxy clusters.

In this paper we describe a theory of CRs which combines a given radio halo detection with a $\gamma$-ray flux upper limit to derive a minimum CRp transport speed. In simple terms, if CRs are being accelerated in a cluster at a rate consistent with its radio emission, they must escape the cluster on short enough time scales in order to bring the CRp density low enough to explain the lack of $\gamma$-rays. This escape rate depends on the assumed parameters of our model, including magnetic field topology and relative acceleration efficiency of protons to electrons. We will specify to the Coma cluster, but the analysis is general.

It should be noted that ours is far from the first study of cosmic ray transport in galaxy clusters. \pagebreak  \cite{ensslin11} assumed radial transport at the local sound speed. \cite{wiener13a} and \cite{wiener18} assumed radial transport at a speed determined by equating the rate at which cosmic rays excite the waves that confine them through the streaming instability to the rate at which the waves are damped in the cluster plasma (see \cite{zweibel17} for a review). These studies are appropriate when the magnetic field is relatively well ordered and the cosmic rays stream down their density gradient at all times. Other work has, in contrast, suggested that CR diffusion times are of order the age of the universe, and so CRs are effectively confined to their host galaxy clusters (\cite{volk96}, \cite{berezinsky97}, \cite{brunetti14}). This can occur if there are enough externally-driven MHD waves that the self-confinement picture fails, or if the magnetic field is tangled enough that even CRs free streaming at the speed of light do not leave the cluster in a Hubble time. Here we do not ascribe to any particular magnetic field topology or level of turbulence, but instead seek to translate observations of radio and $\gamma$-ray emission into constraints on a set of cluster parameters which includes them. Additional observational constraints on the magnetic field in a cluster would then serve, combined with this analysis, to constrain the remaining model parameters, and vice versa.

\begin{figure*}
\centering
\includegraphics[width=0.32\textwidth]{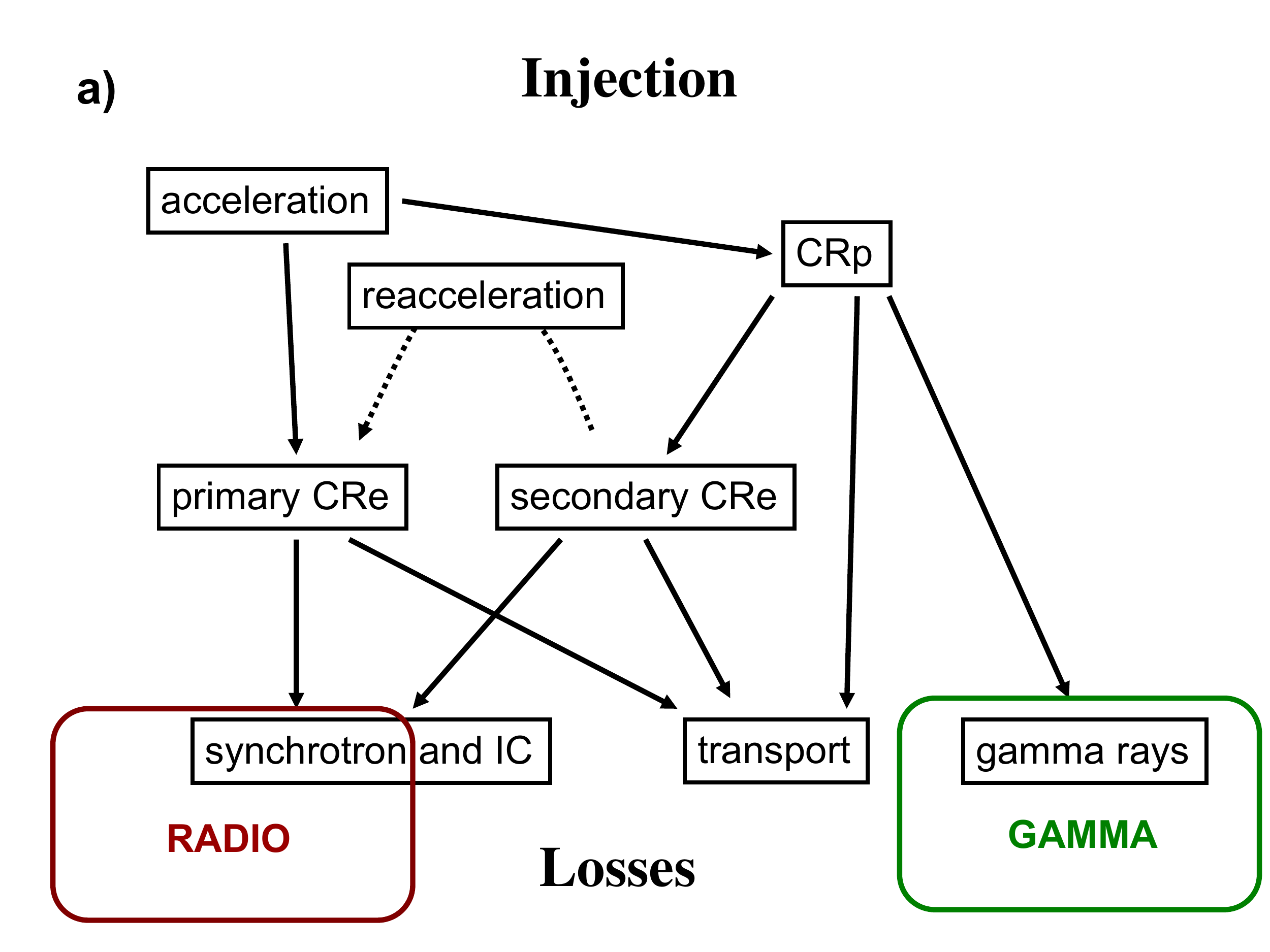}
\includegraphics[width=0.32\textwidth]{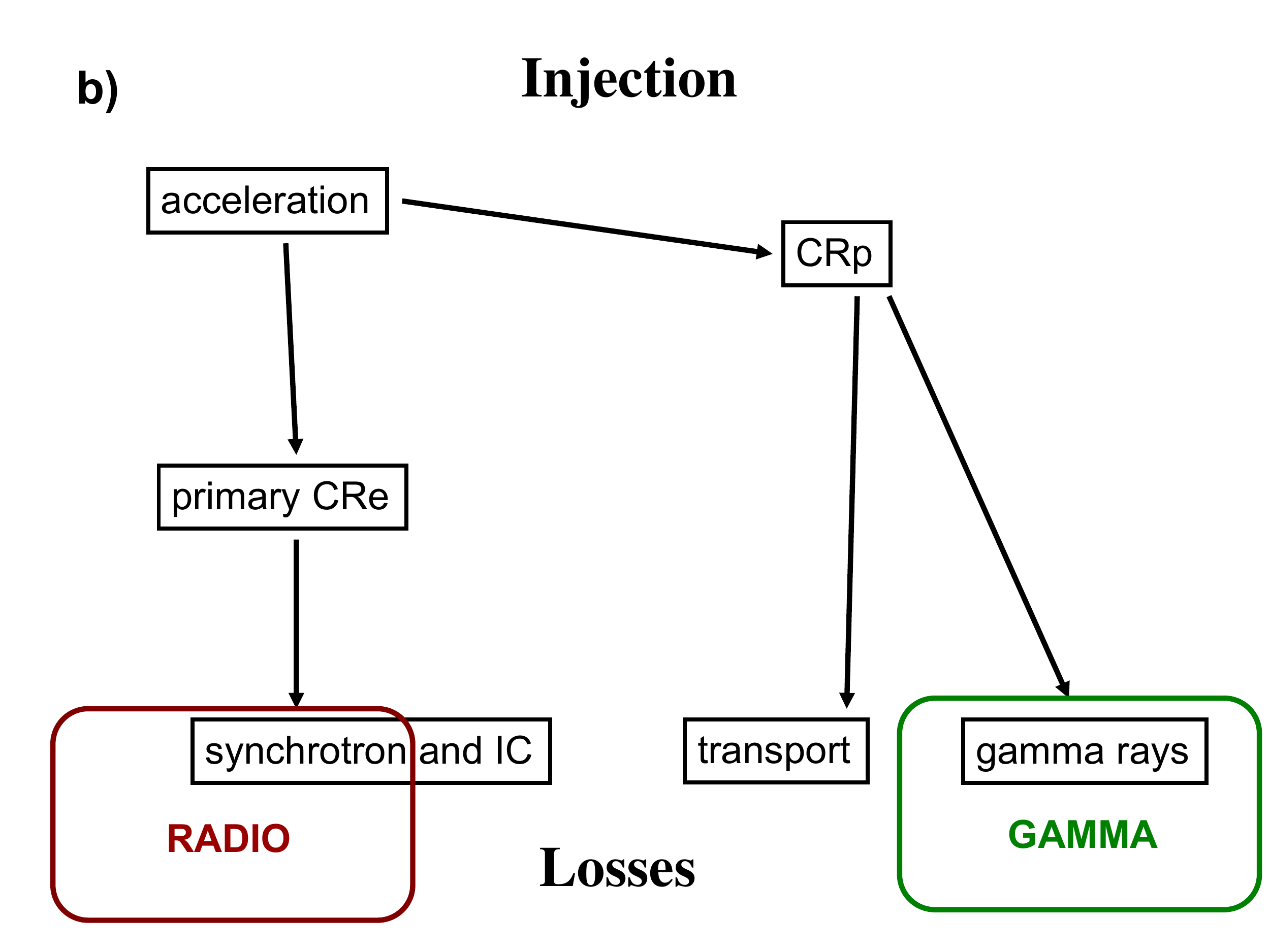}
\includegraphics[width=0.32\textwidth]{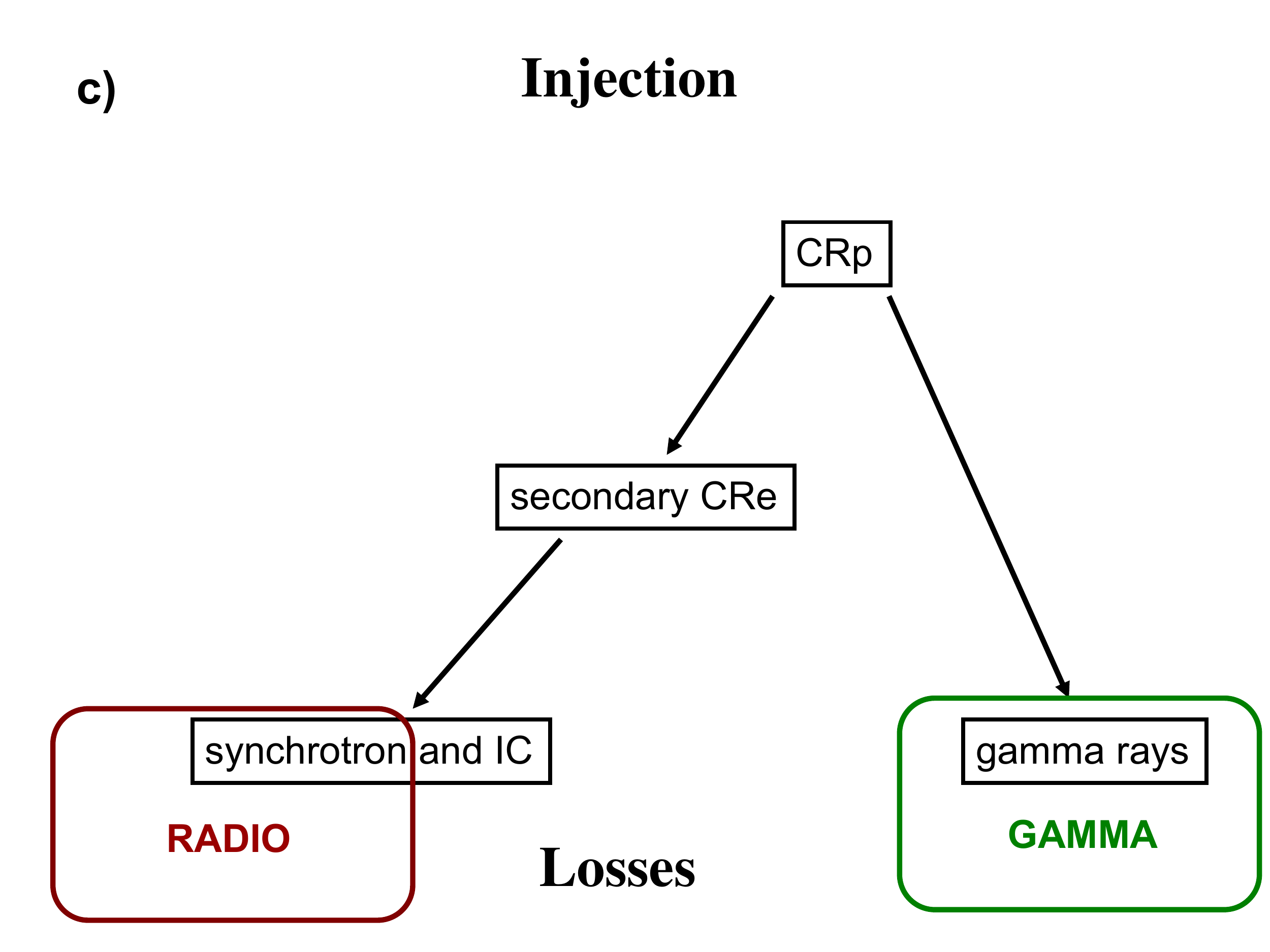}
\caption{Schematic of relevant processes which dictate the density of CRs in a galaxy cluster and the resulting observable radio and $\gamma$-radiation. Left: The full picture with all dominant processes. Middle: The assumption that CRe primaries dominate. Right: The assumption that CRe secondaries dominate.}\label{fig:schematics}
\end{figure*}

In \S\ref{sec:transport} we briefly review the theory of CR transport in the regime relevant to galaxy clusters. We establish a framework for discussing the evolution and steady state conditions of CR electrons and protons, and make some limiting assumptions. We broadly consider two transport models, bulk advection or streaming at speed $v_t$, and diffusion with diffusivity $\kappa$.

In \S\ref{sec:radio} we review the connection between synchrotron emission and the underlying CRe population. From the implied CRe production rate we determine the CRp population in two limits. In \S\ref{sec:primlim}, we assume that the CRe are all primaries (i.e. accelerated by the same processes that produce CRp). In this limit, which is akin to the re-acceleration model of radio halos, the CRp production rate is determined, and we must model CRp transport in order to predict an equilibrium CRp density to compare against $\gamma$-ray observations. In \S\ref{sec:seclim}, we assume instead that the CRe are all secondary particles, i.e. the hadronic model of radio halos. In this limit the predicted CRe production rate translates directly into a CRp density and no constraints on CRp transport are made.

In \S\ref{sec:gamma} we review the connection between $\gamma$-ray emission and CRp to derive constraints on the CRp population from the upper limits on the $\gamma$-ray flux. We combine the analysis of \S\ref{sec:radio} to derive (parameter-dependent) lower limits on CR transport in the case where CRe are dominated by primaries. This is pictorialized in the simplified schematic of CR processes shown in figure \ref{fig:schematics}b. The opposite case, where CRe are dominated by secondaries, is pictorialized in the simplified schematic shown in figure \ref{fig:schematics}c. In this case, rather than limits on CR transport, we get predictions of $\gamma$-ray emission to compare directly with observations. In \S\ref{sec:results} we apply the analysis in \S\S\ref{sec:radio} and \ref{sec:gamma} to the Coma cluster. In \S\ref{sec:checks} we check some of the assumptions made in the analysis for consistency. We find that the minimum transport rates derived for CRp in the primary model are so slow that appreciable secondaries would be generated before the CRp escape.

In \S \ref{sec:conc} we summarize and draw conclusions. Our main result is that if the cosmic ray electrons are primary particles, then with standard assumptions about the magnetic field strength and spatial profile in Coma, and Milky Way like assumptions about the ratio of proton to electron cosmic ray sources, the streaming speeds calculated in \cite{wiener13a} and \cite{wiener18} are high enough to prevent proton cosmic rays from generating a detectable $\gamma$-ray signal provided the magnetic field is not tangled below about 100 kpc. But, if the magnetic field is dominated by turbulence on 10s of kpc scales, as suggested by the observations of \cite{bonafede10}, cosmic rays build up in intensity to a level that should produce a detectable $\gamma$-ray signal for certain choices of our model parameters. The $\gamma$-ray non-detections therefore impose limits on these parameters. One potential use of this limit is to constrain the relative acceleration efficiency of protons to electrons in a cluster. In the opposite limit where the cosmic ray electrons are all secondaries, we find that the radio detections and $\gamma$-ray non-detections require a rather strong magnetic field.

\section{Cosmic Ray Transport}\label{sec:transport}
In this section we introduce the basic CR transport equations. We first set up general evolution equations that we will use in our analysis later on. These are general in the sense that they encompass a large family of specific evolution models. We follow this with a brief description of one of these specific models, the self-confinement regime of CR transport. See \cite{zweibel17}, \cite{skilling71}, \cite{longair92}, or \cite{kulsrud05} for some examples of further review of CR transport equations beyond what is covered here.

\subsection{A Generalized Treatment}\label{subsec:general}
The population of cosmic rays in a system can be completely described by its phase-space density function $f_i(\mb{r},\mb{p})$, which is defined by
\begin{equation}
\tr{d}^6N_i(\mb{r},\mb{p})=f_i(\mb{r},\mb{p})\tr{d}^3\mb{r}\tr{d}^3\mb{p},
\end{equation}
where d$^6N_i$ indicates the number of CRs in the phase-space volume $\tr{d}^3\mb{r}\tr{d}^3\mb{p}$ centered at $(\mb{r},\mb{p})$. The subscript $i$ indicates different CR species ($e$ for electrons, $p$ for protons, etc.). The evolution of such a CR population is then most generally described by the Vlasov equation, which expresses the conservation of particles in phase-space:
\begin{equation}\label{eq:vlasov}
\pp{f_i(\mb{r},\mb{p})}{t}+\nabla\cdot(f_i\mb{v})+\nabla_\mb{p}\cdot\left(f_i\dd{\mb{p}}{t}\right)=0.
\end{equation}
(As written, equation (\ref{eq:vlasov}) does not account for injection of new particles; we add a source term below, in equation\eqref{eq:transporteq}.)

We simplify equation (\ref{eq:vlasov}) by imposing the limits inherent in our system. We will approximate our galaxy clusters as spherically symmetric systems, so we can integrate \eqref{eq:vlasov} over the two angular coordinates. Furthermore, in the limit of small mean free path due to scattering, CRs are nearly isotropic in momentum space. Equation \eqref{eq:vlasov} then reduces to
\begin{multline}\label{eq:transporteq}
\pp{f_i(r,E_i)}{t}+\frac{1}{r^2}\pp{}{r}(r^2F_{cr,i}(r,E_i))\\+\pp{}{E_i}\left(\dot{E}_i(r,E_i)f_i(r,E_i)\right)
=s_i(r,E_i),
\end{multline}
where the reduced distribution function is now defined by
\[
\tr{d}^2N_i(r,E_i)=4\pi r^2f_i(r,E_i)\tr{d}r\tr{d}E_i.
\]
In the above, $\tr{d}^2N_i$ represents the number of CRs of species $i$ between cluster radii $r$ and $r+\tr{d}r$ and having energy between $E$ and $E+\tr{d}E$. In \eqref{eq:transporteq}, $F_{cr,i}$ is the spherically-averaged CR number flux per unit energy due to transport, which will be determined by the given CR transport model. Changes in energy are represented by the third term, with $\dot{E_i}$ being the rate of energy loss or gain per CR. Processes which produce CRs, such as acceleration or production in hadronic collisions, are incorporated in the source function $s_i$, which has dimensions of number per volume per unit energy per unit time.

The main goal of this paper will be to use cluster observations to constrain $F_{cr,i}$, and so we do not restrict the analysis to any specific CR transport model. In \S\ref{subsec:specific} we consider the self-confinement model of transport for illustrative purposes and to connect the general transport framework to a physical theory, but the analysis presented in \S\S\ref{sec:radio} and \ref{sec:gamma} is intended to be as general as possible.

We next apply our limiting assumptions for CRe and CRp respectively to \eqref{eq:transporteq} and derive steady state equations for each. Steady state should be a reasonable approximation for CRe, which have short cooling times of $\approx100$ Myr for the relevant energies of a few GeV. If the CRe sources in the galaxy cluster do not change faster than this time scale, the CRe density will be held in steady state. Even if the sources fluctuate on faster time scales, the radiative losses would respond to an effective time-smoothing of the CRe source. The resulting CRe density would be a steady state between the radiative cooling and this time-averaged source. The ``sources" used below should then be understood to be time-smoothed on 100 Myr scales. 

The steady state assumption is less robust for CRp which have much longer cooling times. CRp sources are then primarily offset by transport, which may in principle be very slow. The current CRp density in a region of size $L$ is then determined by the CRp injection history within that region over a CRp crossing time. Assuming the CRp are in steady state would therefore overpredict (underpredict) the real density of CRp if the rate of CRp injection is higher (lower) than the average over this time period. Again, any sources discussed below can be thought of as time-smoothed versions of some real CRp source. Further consideration of time dependent models is beyond the scope of this paper.

As stated above, CRe suffer significant energy losses to synchrotron radiation and inverse Compton (IC) emission. We will therefore approximate advection and diffusion losses as negligible, $\mb{F_{cr,e}}=0$, an approximation we will check after the fact in \S\ref{sec:checks}. In the steady state, \eqref{eq:transporteq} then reduces to
\begin{equation}
\pp{}{E_e}\left(\dot{E}_e(r,E_e)f_e(r,E_e)\right)
=s_e(r,E_e).
\end{equation}
This has the simple interpretation that, in the steady state, the divergence of any flux of CRe in energy space must be balanced by a CRe source at every point. A CRe distribution function $f_e(r,E_e)$ that suffers energy losses $\dot{E}_e<0$ and is replenished by source function $s_e(r,E_e)$ therefore has the steady state solution
\begin{equation}\label{eq:steadye}
f_e(r,E_e)=\frac{1}{|\dot{E}_e|}\int_{E_e}^\infty s_e(r,E'_e)\tr{d}E'_e.
\end{equation}

It will be helpful to work in terms of Lorentz factor $\gamma_e=E_e/m_ec^2$ instead of $E_e$. We define $\gamma_e$-dependent functions $f_e$ and $s_e$ as
\[
f_e(r,\gamma_e)=f_e(r,E_e)\dd{E_e}{\gamma_e}=m_ec^2f_e(r,E_e),
\]
\[
s_e(r,\gamma_e)=m_ec^2s_e(r,E_e).
\]
Equation \eqref{eq:steadye} is then written
\begin{equation}\label{eq:steadyegamma}
f_e(r,\gamma_e)=\frac{1}{|\dot{\gamma}_e|}\int_{\gamma_e}^\infty s_e(r,\gamma'_e)\tr{d}\gamma'_e.
\end{equation}

CR protons above about 1 GeV have much longer cooling times than electrons of the same energy (see \cite{ensslin11} Figure 2). For protons we therefore approximate the energy losses as negligible, $\dot{E}_p=0$ (Adiabatic losses, which are also included in $\dot{E}_p$, may be more significant. We discuss these losses for the specific case of CR self-confinement in \S\ref{subsec:specific} and the consider the effect of ignoring them in our analysis in \S\ref{sec:checks}). Equation \eqref{eq:transporteq} then reduces, in the steady state, to
\begin{equation}\label{eq:fluxtrans}
\frac{1}{r^2}\pp{}{r}(r^2F_{cr,p}(r,E_p))=s_p(r,E_p).
\end{equation}
Again this has a simple interpretation that in the steady state, any CRp which leave a given volume must be replenished by the source of CRp in that volume. The solution, assuming the net transport is outward ($F_{cr,p}>0$), is
\begin{equation}\label{eq:steadyp}
F_{cr,p}=\frac{1}{r^2}\int_0^r r'^2s_p(r',E_p)\tr{d}r'.
\end{equation}

To get any further we need to define $F_{cr,p}$, i.e. we need categorize a model of CRp transport. We consider two classes of models here, advection and diffusion. We reiterate that we do not limit analysis to any specific model of CR transport in this work. We use advection and diffusion as two common limiting cases of CR transport that encompass several specific transport models.

We first consider advection, wherein we characterize radial transport by some bulk speed $v_t$ which may in principle vary with position and CRp energy:
\begin{equation}\label{eq:fluxadv}
F_{cr,p}(r,E_p)\equiv f_p(r,E_p)v_t(r,E_p).
\end{equation}
As we discuss in \S\ref{subsec:specific}, the self-confinement streaming picture is one example of a CR transport model which has this form. We emphasize that this bulk speed $v_t$ is not the speed of any individual CR, which is close to the speed of light, but rather is a measure of the net flow \emph{defined} by \eqref{eq:fluxadv}.

For a CR flux of this form, the steady state solution is simply
\begin{equation}\label{eq:steadypadv}
f_p(r,E_p)=\frac{1}{r^2v_t(r,E_p)}\int_0^rr'^2s_p(r',E_p)\tr{d}r'.
\end{equation}
Evaluating this requires knowing $v_t$ and $s_p$. This is done in \S\ref{sec:primlim}.

We next consider a diffusive model for CR transport, where the CRp flux is
\begin{equation}\label{eq:fluxdiff}
F_{cr,p}(r,E_p)\equiv-\kappa(r,E_p) \pp{f_p(r,E_p)}{r}
\end{equation}
for some diffusion coefficient $\kappa$. In this regime, the steady state requirement \eqref{eq:steadyp} becomes
\[
\pp{f_p}{r}=-\frac{1}{r^2\kappa(r,E_p)}\int_0^r r'^2 s_p(r',E_p)\tr{d}r'.
\]
We can find $f_p(r,E_p)$ from this by assuming the CRp density at some outermost cluster radius $r_\tr{max}$ is negligible and integrating inward:
\begin{equation}\label{eq:steadypdiff}
f_p(r,E_p)=\int_r^{r_\tr{max}}\frac{\tr{d}r''}{(r'')^2\kappa(r'',E_p)}\int_0^{r''} r'^2 s_p(r',E_p)\tr{d}r'.
\end{equation}
We evaluate this expression in \S\ref{sec:primlim}.

\subsection{A Specific Treatment}\label{subsec:specific}
In this section we consider a specific model of CR transport and show how it amounts to one case of the general framework of \S\ref{subsec:general}. We describe the evolution of CRs in the specific context where transport is mediated by scattering off of Alfv\'en waves traveling in both directions along the background magnetic field. The self-confinement streaming picture is a special case of this. In the self-confinement model, the majority of Alfv\'en waves present are generated by CR pressure gradients via the CR streaming instability. CRs will then flow in bulk at a speed which is determined by marginal stability of these waves, and waves will only travel in one direction (see \cite{skilling71}).

\cite{skilling75a} lays out the derivation in detail: starting from the Vlasov equation \eqref{eq:vlasov}, we first assume the fastest timescale in the problem is the cosmic ray gyromotion in the background magnetic field, and average \eqref{eq:vlasov} over gyration angle. We next assume the second fastest timescale is the frequency of scattering from Alfv\'en waves traveling in both directions along the magnetic field. Denote the scattering rate of CRs off of forward- backward-moving Alfv\'en waves as $\nu_\pm$, which will depend on the CR momentum $p$ and pitch angle cosine $\mu$. \cite{skilling75a} defines a (momentum-dependent) wave frame
\begin{equation}
\mb{u}=\mb{v_0}+\left\langle\frac{3}{2}(1-\mu^2)\frac{\nu_+-\nu_-}{\nu_++\nu_-}\right\rangle v_A\mb{n}
\end{equation}
where $\mb{v_0}$ is the velocity of the gas, $v_A=B/\sqrt{4\pi \rho}$ is the Alfv\'en speed, and $\mb{n}$ is a unit vector pointing in the direction of the background field. The angle brackets indicate an average over $\mu$. Note that in the limit where waves only travel in one direction, as would be expected in the self-confinement regime, this frame reduces to $\mb{u}=\mb{v_0}\pm v_A\mb{n}$ for all momenta. Strictly, the self confinement model describes transport along magnetic fieldlines, which are not necessarily radial, and predicts its own breakdown in situations where the field aligned components of the thermal gas density gradient and cosmic ray pressure gradient are antiparallel \citep{skilling71, wiener17a}, which is one reason why we consider more general models here. 

Shifting the gyration-averaged equation into this frame and averaging over pitch angle, \cite{skilling75a} arrives at the following equation for $\bar{f}_i(\mb{r},p)$:
\begin{multline}\label{eq:transportstreaming}
\pp{\bar{f}_i}{t}+\pp{}{p^3}(p^3\mb{u})\cdot\nabla \bar{f}_i=(\nabla\cdot\mb{u})p^3\pp{\bar{f}_i}{p^3}+\\
\nabla\cdot(D_\parallel\mb{n}\mb{n}\cdot\nabla \bar{f}_i)+\pp{}{p^3}\left(9p^4D_\tr{pp}\pp{\bar{f}_i}{p^3}\right)
\end{multline}
with spatial and momentum diffusion coefficients defined by
\begin{equation}
D_\parallel=v^2\left\langle\frac{1-\mu^2}{2(\nu_++\nu_-)}\right\rangle,
\end{equation}
\begin{equation}
D_\tr{pp}=4\gamma^2m^2v_A^2\left\langle\frac{1-\mu^2}{2}\frac{\nu_+\nu_-}{\nu_++\nu_-}\right\rangle.
\end{equation}
Here, $D_\parallel$ represents diffusion in space due to pitch-angle scattering by the Alfv\'en waves, while $D_\tr{pp}$ represents diffusion in energy due to Fermi acceleration.

Note that $\bar{f}_i(\mb{r},p)$ here is the average of the phase-space density $f_i(\mb{r},\mb{p})$ over all momentum angles:
\[
\bar{f}_i(\mb{r},p)=\frac{1}{4\pi}\iint\tr{d}^2\Omega_p\frac{\tr{d}^6N}{\tr{d}^3\mb{r}\tr{d}^3\mb{p}}=\frac{1}{p^2}\frac{\tr{d}^4N}{\tr{d}^3\mb{r}\tr{d}p}.
\]
This is \emph{not} the same as the energy distribution function $f_i(\mb{r},E)$ introduced in equation \eqref{eq:transporteq} and use throughout \S\ref{subsec:general}, but they are related by\footnote{The angular spatial coordinates have been restored to $f_i$ in this section for ease of comparison to \cite{skilling75a}.}
\begin{equation}
f_i(\mb{r},E)=p^2\bar{f}_i(\mb{r},p)\dd{p}{E}\approx \frac{p^2}{c}\bar{f}_i(\mb{r},p).
\end{equation}
With this in mind, we cast the equation of motion \eqref{eq:transportstreaming} specific to the wave scattering model in the form of \eqref{eq:transporteq}. To do this, note that the second term on the left-hand side of \eqref{eq:transportstreaming} can be written:
\[
\pp{}{p^3}(p^3\mb{u})\cdot\nabla \bar{f}_i=\nabla\cdot\left[\frac{\bar{f}_i}{3p^2}\pp{}{p}(p^3\mb{u})\right]-\frac{\bar{f}_i}{3p^2}\pp{}{p}(p^3\nabla\cdot\mb{u}).
\]
Likewise, the first term on the right-hand side of \eqref{eq:transportstreaming} can be written:
\[
(\nabla\cdot\mb{u})p^3\pp{\bar{f}_i}{p^3}=\frac{1}{p^2}\pp{}{p}\left(\frac{p^3}{3}\bar{f}_i\nabla\cdot\mb{u}\right)-\frac{\bar{f}_i}{3p^2}\pp{}{p}\left(p^3\nabla\cdot\mb{u}\right).
\]
Equation \eqref{eq:transportstreaming} then becomes, after canceling terms and rearranging:
\begin{multline}
\pp{\bar{f}_i}{t}+\nabla\cdot\left[\frac{\bar{f}_i}{3p^2}\pp{}{p}(p^3\mb{u})\right]-\frac{1}{p^2}\pp{}{p}\left(\frac{p^3}{3}\bar{f}_i\nabla\cdot\mb{u}\right)=\\
\nabla\cdot(D_\parallel\mb{n}\mb{n}\cdot\nabla \bar{f}_i)+\frac{1}{p^2}\pp{}{p}\left(p^2D_\tr{pp}\pp{\bar{f}_i}{p}\right),
\end{multline}
or, multiplying both sides by $p^2/c$ and further rearranging terms,
\begin{multline}
\pp{f_i}{t}+\nabla\cdot\left[\frac{f_i}{3E^2}\pp{}{E}(E^3\mb{u})-D_\parallel\mb{n}\mb{n}\cdot\nabla f_i\right]\\-\pp{}{E}\left[\frac{E}{3}f_i\nabla\cdot\mb{u}+D_\tr{pp}\pp{f_i}{E}\right]=0.
\end{multline}
This is now of the same form as \eqref{eq:transporteq}, but without the source term and with an additional diffusion term in energy space. The two terms under the spatial divergence operator nicely resemble the advective and diffusion forms of transport flux respectively,
\[
F_{cr,\tr{adv}}=\frac{f_i}{3E^2}\pp{}{E}(E^3\mb{u})\equiv f_i\mb{v}_{t,\tr{stream}}
\]
\[
\Rightarrow \quad \mb{v}_{t,\tr{stream}}(\mb{r},E)=\frac{1}{3E^2}\pp{}{E}(E^3\mb{u}),
\]
\[
F_{cr,\tr{diff}}=-D_\parallel\mb{nn}\cdot\nabla f_i\equiv -\mb{\sf K}_\tr{stream}(\mb{r},E)\cdot\nabla f_i
\]
\[
\Rightarrow \quad \mb{\sf K}_\tr{stream}(\mb{r},E)=D_\parallel\mb{nn}.
\]
This model of CR transport is one example the generalized treatment of \S\ref{subsec:general}, where the CR flux has an advective piece and a diffusive piece. The energy gains/losses include a Fermi acceleration term and adiabatic energy changes
\begin{equation}\label{eq:adiabatic}
\dot{E}_\tr{adia}=\frac{E}{3}\nabla\cdot\mb{u}
\end{equation}
associated with the advective transport. This is the expected adiabatic change for particles of a fluid with adiabatic index $\gamma=4/3$ and is likely to be a loss term due to the decrease in gas density with radius. It is straightforward to add to this specific model additional loss processes, like synchrotron radiation, or source terms.

The comparison is even clearer in the self-confinement limit, where waves only travel in one direction. Then $D_\tr{pp}=0$ and $\mb{u}=\mb{v_0}\pm v_A\mb{n}$ is independent of $E$, giving:
\begin{equation}
\pp{f_i}{t}+\nabla\cdot\left[f_i\mb{u}-D_\parallel\mb{n}\mb{n}\cdot\nabla f_i\right]-\pp{}{E}\left[\frac{E}{3}f_i\nabla\cdot\mb{u}\right]=0,
\end{equation}
which is just equation \eqref{eq:transporteq} for the specific case of
\[
F_{cr,i}=f_i\mb{u}-D_\parallel\mb{n}\mb{n}\cdot\nabla f_i,
\]
\[
\dot{E}_i=-\frac{E}{3}\nabla\cdot\mb{u}.
\]
This reflects the physical description of the self-confinement picture: CRs are carried along with the Alfv\'en waves, with some amount of diffusion, and suffer adiabatic losses.

\section{Properties Determined from Radio Emission}\label{sec:radio}
\subsection{CRe Profile}\label{subsec:crep}
As per \cite{rybicki79}, an isotropic, spherically symmetric, power law distribution of electrons
\[
f_e(r,\gamma_e)=C_e(r)\gamma_e^{-\alpha_e}
\]
in the presence of a magnetic field $B(r)$ will emit synchrotron radiation with combined power per unit volume per unit frequency given by\footnote{Although we formally integrate the contribution of each electron over all $E_e$ to obtain this expression, only a certain range of energies will contribute significantly. Namely, there is an exponential cutoff in the differential power emitted at low $E_e$ (the location of this cutoff depends on the frequency $\omega$). This means we don't have to worry about the fact that our assumed CRe distribution formally contains some electrons that violate the assumption in \cite{rybicki79} that $\beta\approx 1$. This also means we only need the CRe spectrum to follow a power law in this energy range to be able to use \eqref{eq:ptot}.}:
\begin{multline}\label{eq:ptot}
P_\tr{tot}(\omega,r)=C_e(r)\frac{\sqrt{3\pi}}{4\pi(\alpha_e+1)}\frac{e^3B(r)}{m_ec^2}
\left(\frac{\omega m_ec}{3eB(r)}\right)^{-(\alpha_e-1)/2}\\
\times\frac{\Gamma\left(\frac{\alpha_e}{4}+\frac{19}{12}\right)\Gamma\left(\frac{\alpha_e}{4}-\frac{1}{12}\right)\Gamma\left(\frac{\alpha_e}{4}+\frac{5}{4}\right)}{\Gamma\left(\frac{\alpha_e}{4}+\frac{7}{4}\right)}
\end{multline}

That is, a power law CRe distribution of index $\alpha_e$ emits a power law radio spectrum of index $s=(\alpha_e-1)/2$. Inverting this relationship, an observed power law radio spectrum of index $s$ implies a power law CRe distribution of index $\alpha_e=2s+1$. Let us rewrite \eqref{eq:ptot} in terms of $s$, and also switch to frequency $\nu$ instead of $\omega$ by multiplying by d$\omega$/d$\nu$=2$\pi$.
\begin{equation}\label{eq:ptot2}
P_\tr{tot}(\nu,r)=\mathcal{N}C_e(r)\frac{e^3B(r)}{m_ec^2}\left(\frac{2\pi\nu m_ec}{3eB(r)}\right)^{-s}
\end{equation}
\[
\mathcal{N}=\frac{\sqrt{3\pi}}{4s+4}\frac{\Gamma\left(\frac{s}{2}+\frac{11}{6}\right)\Gamma\left(\frac{s}{2}+\frac{1}{6}\right)\Gamma\left(\frac{s}{2}+\frac{3}{2}\right)}{\Gamma\left(\frac{s}{2}+2\right)}
\]

To get the observed radio surface brightness at frequency $\nu$ we would take a line integral of $P_\tr{tot}$ along a line of sight through the cluster. This would involve a non-analytic integral at every point on the sky that we wanted to compare to observations. Let us instead consider the total radio power emitted by the entire cluster. Denote the surface brightness as a function of position on the sky (an observed quantity with units Jy sr$^{-1}$) by $S_\nu(\Omega)$. The power per unit volume per frequency bin in radio (an intrinsic quantity with units erg cm$^{-3}$ s$^{-1}$ Hz$^{-1}$) is $P_\tr{tot}(\nu,r)$ as given above in \eqref{eq:ptot2}. Then the total observed flux can be expressed in two ways:
\[
\int\tr{d}^2\Omega S_\nu(\Omega)=\frac{1}{4\pi D^2}\int\tr{d}^3VP_\tr{tot}(\nu,r)
\]
\begin{equation}\label{eq:totint}
\int_0^{\theta_\tr{max}}2\pi\sin\theta\tr{d}\theta S_\nu(\theta)=\int_0^{r_\tr{max}}\tr{d}r\frac{r^2}{D^2}P_\tr{tot}(\nu,r)
\end{equation}
Here we have assumed the cluster is spherically symmetric, is a distance $D$ away, and has some maximum extent on the sky $\theta_\tr{max}=r_\tr{max}/D$. We have also neglected geometric effects of the size of the cluster by approximating the distance to every point in the cluster as $D$. The left hand side of this equation encases the observations, while the right hand side, via equation \eqref{eq:ptot2}, contains parts of our cluster model $B(r)$ and $C_e(r)$.

Let us plug in \eqref{eq:ptot2} for $P_\tr{tot}$ in \eqref{eq:totint}, scaling to the central magnetic field $B_0$ and central value of $C_e(r)$, denoted $C_{e0}$. We have
\[
\int_0^{\theta_\tr{max}}2\pi\sin\theta\tr{d}\theta S_\nu(\theta)=
\]
\[
\mathcal{N}C_{e0}\frac{e^3B_0}{m_ec^2}\left(\frac{2\pi\nu m_ec}{3eB_0}\right)^{-s}\int_0^{r_\tr{max}}\tr{d}r\frac{r^2}{D^2}\frac{C_e(r)}{C_{e0}}\left(\frac{B(r)}{B_0}\right)^{s+1}
\]
Thus, for a given model of the magnetic field $B(r)$ and a model for the spatial dependence of $C_e(r)$, we can use the observations $S_\nu$ to determine the normalization of $C_e(r)$.

Let us do this by defining a CRe shape function
\begin{equation}
\eta_e(r)=C_e(r)/C_{e0},
\end{equation}
with the inherent normalization $\eta_e(0)=1$. Denoting the total integrated radio flux per frequency bin at frequency $\nu$ (the left-hand side of \eqref{eq:totint}) by $S_\nu$ and solving for $C_{e0}$, we arrive at
\begin{equation}\label{eq:ce0}
C_{e0}=\frac{\mathcal{S}}{\mathcal{N}\mathcal{I}_1}\frac{m_ec^2D^2}{e^3B_0}
\end{equation}
where
\begin{equation}
    \mathcal{S}=S_\nu\left(\frac{2\pi\nu m_ec}{3eB_0}\right)^s
\end{equation}
is a single constant that encompasses the radio observations and
\begin{equation}\label{eq:i1}
    \mathcal{I}_1=\int_0^{r_\tr{max}}\tr{d}r r^2\eta_e(r)\left(\frac{B(r)}{B_0}\right)^{s+1}
\end{equation}
is an integral containing information about the cluster model. We motivate some possible forms of $\eta_e(r)$ and $B(r)$ in \S\ref{shape}. As we discuss in \S\ref{sec:results}, our final results are insensitive to our choice of $\eta_e(r)$.

\subsection{CRe Loss and Source Rate}\label{sec:eloss}

In the relevant CRe energy range that is probed by the radio observations as described above, energy losses are dominated by synchrotron and inverse Compton (IC) losses. The photon field responsible for the IC losses is itself dominated by cosmic microwave background (CMB) photons. The energy loss rate per electron is then
\begin{multline}\label{eq:radloss}
\dot{\gamma}_e\tr{(per electron)}=-\frac{4}{3}\frac{\sigma_Tc\gamma_e^2}{m_ec^2}(\varepsilon_B(r)+\varepsilon_\tr{cmb})\\
=-\frac{1}{6\pi}\frac{\sigma_Tc\gamma_e^2}{m_ec^2}(B^2(r)+B^2_\tr{cmb})
\end{multline}
where $B_\tr{cmb}\approx 3.24\ \mu$G is the magnetic field with equivalent energy density as the CMB.

Recall from \S\ref{sec:transport} that the steady state equation for an isotropic, spherically-symmetric distribution of CRe, neglecting transport losses, is given by \eqref{eq:steadyegamma}:
\[
f_e(r,\gamma_e)=\frac{1}{|\dot{\gamma}_e|}\int_{\gamma_e}^\infty s_e(r,\gamma'_e)\tr{d}\gamma'_e.
\]
For our power law distribution, $f_e(r,\gamma_e)=C_{e0}\eta_e(r)\gamma_e^{-\alpha_e}$. Since $\dot{\gamma}_e$ goes as $\gamma_e^2$, the source function's energy dependence must be $\gamma_e^{1-\alpha_e}$. Let us then write the source function in the form $s_e(r,\gamma_e)=G_e(r)\gamma_e^{1-\alpha_e}$. Then
\[
C_{e0}\eta_e(r)\gamma_e^{-\alpha_e}=\frac{6\pi m_ec^2}{\sigma_Tc\gamma_e^2(B^2(r)+B_\tr{cmb}^2)}G_e(r)\int_{\gamma_e}^\infty {\gamma'_e}^{1-\alpha_e}\tr{d}\gamma'_e
\]
\[
C_{e0}\eta_e(r)=\frac{6\pi m_ec^2G_e(r)}{(\alpha_e-2)\sigma_Tc(B^2(r)+B_\tr{cmb}^2)}
\]
Inverting this relationship gives us the source function $G_e(r)$ required to give a steady state CRe distribution $C_{e0}\eta_e(r)$:
\begin{equation}\label{eq:esource0}
G_e(r)=\frac{\alpha_e-2}{6\pi}\frac{\sigma_Tc}{m_ec^2}C_{e0}\eta_e(r)(B^2(r)+B_\tr{cmb}^2)
\end{equation}
If we use our formula for $C_{e0}$ based on the radio emission \eqref{eq:ce0}, we get
\[
s_e(r,\gamma_e)=G_e(r)\gamma_e^{-2s},
\]
\begin{equation}\label{eq:esource}
G_e(r)=\frac{(2s-1)\mathcal{S}}{6\pi\mathcal{N}\mathcal{I}_1}\frac{\sigma_TcD^2}{e^3B_0}\eta_e(r)(B^2(r)+B_\tr{cmb}^2)
\end{equation}
Put simply, for a given cluster model, the spatial shape of the source function (assuming the CRe are in steady state) is uniquely determined, and the radio observations determine its normalization.

\subsection{CRe Shape Function}\label{shape}
Before moving on it will be useful to come up with an educated guess for what the CRe shape function $\eta_e(r)$ should be. To do this, consider the more complicated problem of looking at the radio emission as a function of position. This involves a line of sight integral
\[
S_\nu(\theta)=\frac{1}{2\pi}\int_0^\infty\tr{d}lP_\tr{tot}(\nu,r(l)=\sqrt{D^2\theta^2+l^2})
\]
with $P_\tr{tot}$ given by equation \eqref{eq:ptot2}. 

Recall that the normalization of $\eta_e(r)$ is arbitrary - we are only interested in its spatial dependence. If we just want to find this dependence, we can ignore all factors except those which depend on $r$. We are left with the relationship between the radio brightness $S_\nu(\theta)$ and a line integral of the shape function:
\[
S_\nu(\theta)\propto \int_0^\infty\tr{d}l\ \eta_e(r(l))\left(\frac{B(r(l))}{B_0}\right)^{s+1}.
\]

From here on we will model the Coma cluster specifically, but the analysis can be applied more generally. We model the B field of the Coma cluster as one that goes as a power $\alpha_B$ of the thermal electron density,
\begin{equation}\label{eq:bfield}
B(r)=B_0\left(\frac{n_e(r)}{n_0}\right)^{\alpha_B}.
\end{equation}
Let us consider shape functions of the same form:
\begin{equation}\label{eq:shape1}
\eta_e(r)=\left(\frac{n_e(r)}{n_0}\right)^{\alpha_\eta}
\end{equation}
going as the electron density to some power $\alpha_\eta$. The electron density itself can be modeled with a beta profile:
\begin{equation}\label{eq:beta}
n_e(r)=n_0\left(1+\left(\frac{r}{r_c}\right)^2\right)^{-3\beta/2}
\end{equation}
with core radius $r_c=294$ kpc and $\beta=0.75$ for the specific case of the Coma cluster determined from X-ray surface brightness in \cite{briel92}. The determination of the shape function thus reduces to determining the power $\alpha_\eta$ from
\[
S_\nu(\theta)\propto \int_0^\infty\tr{d}l\ \left(1+\frac{l^2+D^2\theta^2}{r_c^2}\right)^{-(3\beta/2)(\alpha_B(s+1)+\alpha_\eta)}
\]
\[
S_\nu(\theta)\propto \int_0^\infty\tr{d}l\ \left(l^2+r_c^2+D^2\theta^2\right)^{-(3\beta/2)(\alpha_B(s+1)+\alpha_\eta)}
\]
The primary dependence on $\theta$ (at least for large $\theta$) appears to be $\theta^{-3\beta(\alpha_B(s+1)+\alpha_\eta)}$. We can compare this to the $\theta$ dependence of $S_\nu(\theta)$ at large $\theta$ which is about the same as that of the X-ray surface brightness (very roughly speaking - see \cite{deiss97} figure 3), $S_X(\theta)\sim\theta^{1-6\beta}$. We end up with
\[
1-6\beta=-3\beta(\alpha_B(s+1)+\alpha_\eta)
\]
\begin{equation}
\rightarrow \alpha_\eta=2-\frac{1}{3\beta}-\alpha_B(s+1)
\end{equation}
For $\beta=0.75$, $\alpha_B=0.5$, and $s=1.35$, we get $\alpha_\eta=.381$.

Since the extent of the radio halo may depend on frequency and data analysis techniques (\cite{brown11-coma} find a more extended halo in the Coma cluster than \cite{deiss97} even at the same frequency), we will investigate the effects of changing $\alpha_\eta$, using the above formula as a fiducial value. As we explain in \S\ref{sec:results}, while there is some dependence of our results with $\alpha_\eta$ (which we vary between 0 and 1), it is subdominant compared to other dependencies. This suggests our results are not sensitive to the assumed shape function $\eta_e$.

\subsection{Cosmic Ray Proton (CRp) Density}
The connection between CRe and cosmic ray protons (CRp) can be complicated, but we can consider two limiting cases. In one limit, CRe secondaries produced by hadronic interactions are negligible in number compared to CRe primaries directly accelerated by the plasma. In the other limit, direct acceleration is negligible, and the CRe are entirely sourced by hadronic reactions. We consider both limits in turn, and check for consistency in \S\ref{sec:checks}.


\subsubsection{Primary limit}\label{sec:primlim}
In this limit we simply assume some relative acceleration efficiency of protons to electrons $\zeta\approx100$,\footnote{This value is highly uncertain, so we merely use 100 as a reference value. As discussed in \S\ref{sec:primres}, the dependence of our results on this parameter will turn out to be linear, so our exact choice for this value is not important.} and the CRp source function is just our result from the previous section times $\zeta$:
\[
s_p(r,E_p)=\zeta s_e(r,E_e=E_p)
\]
\[
=\frac{\zeta}{m_ec^2}s_e\left(r,\gamma_e=\frac{E_p}{m_ec^2}\right)
\]
\begin{equation}\label{eq:psource}
s_p(r,E_p)=\frac{\zeta}{m_ec^2}G_e(r)\left(\frac{E_p}{m_ec^2}\right)^{-2s}.
\end{equation}

If we want to find a steady state distribution function $f_p(r,E_p)$, we need to balance the above source function with some loss terms. In \S\ref{sec:transport} we found that the steady state equation for CRp, where we ignore energy losses, is given by \eqref{eq:steadyp}. We specialized to two classes of transport - advection with some bulk flow $v_t(r,E_p)$ and diffusion with some coefficient $\kappa(r,E_p)$. For advective transport we derived \eqref{eq:steadypadv}:
\[
f_p(r,E_p)=\frac{1}{r^2v_t(r,E_p)}\int_0^rr'^2s_p(r',E_p)\tr{d}r'.
\]

Supposing the functional form of the transport speed $v_t$ is known, this, combined with \eqref{eq:esource} and \eqref{eq:psource}, gives us a way to determine the CRp distribution from the radio emission. Namely,
\[
f_p(r,E_p)=\frac{\zeta}{r^2v_t(r,E_p) m_ec^2}\left(\frac{E_p}{m_ec^2}\right)^{-2s}\int_0^r {r'}^2 G_e(r')\tr{d}r'
\]
\begin{equation}\label{eq:fsteady1}
f_p(r,E_p)=\frac{\zeta\sigma_TcB_0D^2}{r^2v_te^3m_ec^2}\frac{(2s-1)\mathcal{S}\mathcal{I}_2(r)}{6\pi\mathcal{N}\mathcal{I}_1}\left(\frac{E_p}{m_ec^2}\right)^{-2s}
\end{equation}
where we have defined another moment of the model profiles
\begin{equation}\label{eq:i2}
    \mathcal{I}_2(r)=\int_0^r\tr{d}r'{r'}^2\eta_e(r')\left(\frac{B^2(r')+B^2_\tr{cmb}}{B_0^2}\right).
\end{equation}

We see from \eqref{eq:fsteady1} that the energy dependence of the steady state of $f_p$ goes as $E_p^{-2s}/v_t$, and so depends on the energy dependence of $v_t$. For instance, if we assume our transport speed is independent of energy, then we have a power law CRp distribution\footnote{We use the tilde notation $\tilde{C_p}$ here to highlight the difference in dimensionality between it and $C_e$ from the previous section. This arises from defining our CRe distribution function as a function of Lorentz factor $\gamma_e$ and our CRp distribution function as a function of energy $E_p$.} $f_p(r,E_p)=\tilde{C}_p(r)E_p^{-\alpha_p}$ with spectral index $\alpha_p=2s=\alpha_e-1$ and spatial dependence
\begin{equation}\label{eq:primlim}
\frac{\tilde{C}_p(r)}{(m_ec^2)^{2s-1}}=\frac{\zeta\sigma_TcB_0D^2}{r^2v_t(r)e^3}\frac{(2s-1)\mathcal{S}\mathcal{I}_2(r)}{6\pi\mathcal{N}\mathcal{I}_1}.
\end{equation}
Transport speeds which depend on energy as a power law will have the same form, but with a different exponent $\alpha_p$ that is straightforward to determine. More complicated transport speeds require their own treatment on a case by case basis and will not be considered here. Since we only invoke $f_p$ here at energies between around 10 and 100 GeV, this analysis covers any transport processes that can be reasonably approximated by a power law over this range.

We next consider diffusion. In \S\ref{sec:transport} we found the steady state equation \eqref{eq:steadypdiff} for CRp undergoing diffusive transport:
\[
f_p(r,E_p)=\int_r^{r_\tr{max}}\frac{\tr{d}r''}{(r'')^2\kappa(r'',E_p)}\int_0^{r''} r'^2 s_p(r',E_p)\tr{d}r'
\]
Using what we know for $s_p$ and assuming the previously given power law form for $f_p$ (which is valid when we assume $\kappa$ has no explicit spatial dependence and is a power law of energy), we get
\[
\tilde{C}_p(r)E_p^{-\alpha_p}=\frac{\zeta}{\kappa m_ec^2}\left(\frac{E_p}{m_cc^2}\right)^{-2s}
\]
\[
\times\int_r^{r_\tr{max}}\frac{\tr{d}r''}{(r'')^2}\int_0^{r''} r'^2 G_e(r')\tr{d}r'
\]
For an energy-independent $\kappa$ this implies, as before, $\alpha_p=2s$ and gives us the normalization of $f_p$,
\[
\frac{\tilde{C}_p(r)}{(m_ec^2)^{2s-1}}=\frac{\zeta}{\kappa}\int_r^{r_\tr{max}}\frac{\tr{d}r''}{(r'')^2}\int_0^{r''} r'^2 G_e(r')\tr{d}r'
\]
\begin{equation}\label{eq:crpdiff}
\frac{\tilde{C}_p(r)}{(m_ec^2)^{2s-1}}=\frac{\zeta\sigma_TcB_0D^2}{\kappa e^3}\frac{(2s-1)\mathcal{S}\mathcal{I}_3(r)}{6\pi\mathcal{N}\mathcal{I}_1}
\end{equation}
In the above we have defined a third profile integral
\begin{multline}\label{eq:i3}
\mathcal{I}_3(r)=\int_r^{r_\tr{max}}\frac{\tr{d}r''}{(r'')^2}\mathcal{I}_2(r'')\\
=\int_r^{r_\tr{max}}\frac{\tr{d}r''}{(r'')^2}\int_0^{r''}\tr{d}r'r'^2\eta_e(r')\left(\frac{B^2(r')+B_\tr{cmb}^2}{B_0^2}\right).
\end{multline}
As with our previous discussion, if $\kappa$ is a non-zero power law in energy, we end up with much the same result, but with a different exponent $\alpha_p$.

\subsubsection{Secondary limit}\label{sec:seclim}
In the opposite limit, particle acceleration is negligible and the CRe are entirely sourced by hadronic interactions. Schematically, we start with some distribution of CRp $f_p(r,E_p)$ and some cluster gas density model $n(r)$. From these we determine the pion production rate, and thence the production of CRe via the decay of charged pions.

There are a variety of schemes in the literature for determining pion source functions. The delta-function approximation should work for our purposes - we just approximate that every collision of a single CRp of energy $E_p$ with a thermal nucleus results in $\xi$ pions of energy $K_\pi T_p$ each, where $T_p=E_p-m_pc^2$ is the kinetic energy of the incoming proton. In other words, the pion number source function for each collision is
\begin{equation}\label{eq:pionE}
Q_\pi(E_\pi,E_p)=\xi \delta(E_\pi-K_\pi T_p)
\end{equation}
There are different conventions for choosing the values $K_\pi$ and $\xi$ to match experiments. \cite{kelner06} choose $K_\pi\approx0.17$ with $\xi=1$ when calculating the neutral pion source only. \cite{pfrommer08b} use $K_\pi=0.25$ and $\xi=2$ for all pion species together. To be unambiguous, we use $\xi$ here to refer to the total pion multiplicity across all pion species and assume that charged and neutral pions are produced approximately in the ratio 2:1.

If we have a CRp distribution $f_p(r,E_p)$, then the rate of collisions (per unit volume) is $R=cn_N(r)\sigma_{pp}(E_p)f_p(r,E_p)$ and so the pion source function is
\[
q_\pi(r,E_\pi)=\int_0^\infty RQ_\pi(E_\pi,E_p)\tr{d}E_p
\]
\[
=\frac{\xi}{K_\pi} cn_N(r)\sigma_{pp}\left(E_p=\frac{E_\pi}{K_\pi}+m_pc^2\right)f_p\left(r,E_p=\frac{E_\pi}{K_\pi}+m_pc^2\right)
\]
Here, $n_N=n_\tr{H}+4n_\tr{He}$ is the target nucleon density and $\sigma_{pp}$ is the (energy-dependent) cross section for proton-proton collisions. The above expression gives the number source function for all pion types, so by the assumed symmetry the charged pion source function is $s_{\pi^\pm}=2q_\pi/3$.

Let us assume a power law CRp distribution $f_p(r,E_p)=\tilde{C}_p(r)E_p^{-\alpha_p}$. In addition, we will omit the energy dependence of the cross-section $\sigma_{pp}$ with the understanding that it is a very weak function of energy in the relevant energy range around $E_p\sim 100$ GeV (see \cite{kelner06}; \cite{pfrommer08b} account for this weak dependence with an effective cross section that is a function of the CRp power spectrum, $\sigma_{pp}(\alpha_p)=32(0.96+e^{4.4-2.4\alpha_p})$ mbarn). The charged pion source function is then
\begin{multline}
s_{\pi^\pm}(r,E_\pi)=\frac{2}{3}\frac{\xi}{K_\pi}cn_N(r)\sigma_{pp}\\
\times\tilde{C}_p(r)(E_\pi/K_\pi+m_pc^2)^{-\alpha_p}.
\end{multline}

For the next step, we must describe the decay of a charged pion of energy $E_\pi$ into electrons and positrons. Following \cite{pfrommer08b}, we use a delta-function approximation and assume that every charged pion of energy $E_\pi$ decays into exactly one electron/positron of energy $E_e=E_\pi/4$. This means
\[
s_e(r,E_e)=s_{\pi^\pm}(r,E_\pi=4E_e)\dd{E_\pi}{E_e}
\]
\[
=\frac{8}{3}\frac{\xi}{K_\pi}cn_N(r)\sigma_{pp}\tilde{C}_p(r)(4E_e/K_\pi+m_pc^2)^{-\alpha_p}
\]
\begin{equation}\label{eq:secsrc}
s_e(r,E_e)=\frac{8}{3}\frac{\xi}{K_\pi}cn_N(r)\sigma_{pp}\tilde{C}_p(r)(4E_e/K_\pi)^{-\alpha_p}
\end{equation}
In the last step we have assumed we are looking at high energy CRe with $E_e\gg K_\pi m_pc^2/4$. We can also write this in terms of the Lorentz factor $\gamma_e$:
\[
s_e(r,\gamma_e)=\frac{8}{3}\frac{\xi}{K_\pi}cn_N(r)\sigma_{pp}\tilde{C}_p(r)(4\gamma_em_ec^2/K_\pi)^{-\alpha_p}\dd{E_e}{\gamma_e}
\]
\[
=\frac{8}{3}\frac{\xi}{K_\pi}\left(\frac{K_\pi}{4}\right)^{\alpha_p}(m_ec^2)^{1-\alpha_p}cn_N(r)\sigma_{pp}\tilde{C}_p(r)\gamma_e^{-\alpha_p}
\]

We know from \S\ref{sec:eloss} that the CRe source function must be \eqref{eq:esource} to explain the synchrotron emission. In the limit where CRe are completely sourced by hadronic interactions, we therefore have
\[
\frac{2\xi}{3}\left(\frac{4m_ec^2}{K_\pi}\right)^{1-2s}cn_N(r)\sigma_{pp}\tilde{C}_p(r)
\]
\[
=\frac{(2s-1)\mathcal{S}}{6\pi\mathcal{N}\mathcal{I}_1}\frac{\sigma_TcD^2}{e^3B_0}\eta_e(r)(B^2(r)+B_\tr{cmb}^2)
\]
or
\begin{multline}\label{eq:seclim}
\frac{\tilde{C}_p(r)}{(m_ec^2)^{2s-1}}=\frac{(2s-1)\mathcal{S}}{4\xi\pi\mathcal{N}\mathcal{I}_1}\left(\frac{4}{K_\pi}\right)^{2s-1}
\\
\times\frac{\sigma_T}{\sigma_{pp}}\frac{B_0D^2}{e^3}\frac{\eta_e(r)}{n_N(r)}\frac{B^2(r)+B_\tr{cmb}^2}{B_0^2}
\end{multline}

Recall that in the primary limit, we needed, in addition to the radio observations and a cluster model, a model for the transport (in the form of speed $v_t$ or diffusion coefficient $\kappa$) and a relative acceleration efficiency $\zeta$ to determine the spatial distribution of CRp. In the secondary limit, the radio observations and cluster model alone determine the CRp distribution.

\section{Properties Determined from $\gamma$-Ray Emission}\label{sec:gamma}
No diffuse $\gamma$-ray emission has yet been detected from galaxy clusters. At high enough energies, any potential such $\gamma$-rays are expected to be dominated by neutral pion decay. As such, the non-detection of $\gamma$-rays translates to an upper limit on the CRp density. We quantify this below, and follow with a discussion of the meaning of this upper limit in the two limiting cases (primary-dominated and secondary-dominated CRe).

\subsection{CRp Density Upper Limit}\label{sec:gammalim}
The source function of $\gamma$-rays coming from neutral pion decay is a straightforward function of the pion source function:
\begin{equation}
s_\gamma(E_\gamma)=2\int_{E_{\pi,\tr{min}}}^\infty \frac{s_{\pi^0}(E_\pi)\tr{d}E_\pi}{\sqrt{E_\pi^2-m_\pi^2c^4}}
\end{equation}
with $E_{\pi,\tr{min}}=E_\gamma+m_\pi^2 c^4/(4E_\gamma)$. In the previous section we discussed the relationships between the CRp distribution $f_p$ and the pion source function $q_\pi$, as well as between $q_\pi$ and the secondary CRe source function $s_e$.

The neutral pion source function is $s_{\pi^0}=q_\pi/3$. In the previous section we calculated $q_\pi$ for a power law CRp distribution $f_p(r,E_p)=\tilde{C}_p(r)E_p^{-\alpha_p}$. Plugging this into the $\gamma$-ray equation above, we have 
\[
s_\gamma(r,E_\gamma)=\frac{2}{3}\frac{\xi}{K_\pi}cn_N(r)\sigma_{pp}\tilde{C}_p(r)
\]
\[
\times\int_{E_{\pi,\tr{min}}}^\infty\frac{(E_\pi/K_\pi+m_pc^2)^{-\alpha_p}\tr{d}E_\pi}{\sqrt{E_\pi^2-m_\pi^2c^4}}
\]

The above integral cannot be done analytically without some further assumptions. If we assume $E_p=E_\pi/K_\pi\gg m_pc^2$, we can throw away the $m_pc^2$ in the numerator and the integral can be done as an incomplete Beta function. However, in this limit we also have $E_\pi\gg m_\pi c^2$ so the denominator would also simplify, giving us a simple power law. In this limit we also have $E_{\pi,\tr{min}}\approx E_\gamma$. The integral therefore reduces to
\[
K_\pi^{\alpha_p}\int_{E_\gamma}^\infty E_\pi^{-\alpha_p-1}\tr{d}E_\pi=\frac{(E_\gamma/K_\pi)^{-\alpha_p}}{\alpha_p}
\]
and so the $\gamma$-ray source function (just from neutral pion decay) is
\begin{equation}
s_\gamma(r,E_\gamma)=\frac{2\xi}{3\alpha_p}K_\pi^{\alpha_p-1}cn_N(r)\sigma_{pp}\tilde{C}_p(r)E_\gamma^{-\alpha_p}
\end{equation}
and the predicted differential $\gamma$-ray number flux per energy bin at Earth is just the volume integral of this quantity divided by $4\pi D^2$. To connect to a potential observation we would then integrate above some energy $E_{\gamma}$:
\[
F_{\gamma}(>E_{\gamma})=\int_{E_{\gamma}}^{\infty}\tr{d}E_\gamma\int_0^{r_\tr{max}}\tr{d}r \frac{r^2}{D^2}s_\gamma(r,E_\gamma)
\]
\[
=\frac{2\xi K_\pi^{\alpha_p-1}c\sigma_{pp}}{3\alpha_p D^2}\frac{E_{\gamma}^{1-\alpha_p}}{\alpha_p-1}\int_0^{r_\tr{max}}\tr{d}r r^2n_N(r)\tilde{C}_p(r)
\]
\[
=\frac{2\xi K_\pi^{\alpha_p-1}c\sigma_{pp}}{3\alpha_p(\alpha_p-1) D^2}\left(\frac{E_{\gamma}}{m_ec^2}\right)^{1-\alpha_p}\int_0^{r_\tr{max}}\tr{d}r r^2n_N(r)\frac{\tilde{C}_p(r)}{(m_ec^2)^{\alpha_p-1}}
\]

We can now relate a photon number flux upper limit $F_{\gamma,\tr{max}}$ to an upper limit on an integral of the CRp distribution:
\[
F_{\gamma,\tr{max}}(>E_\gamma)\geq \frac{2\xi K_\pi^{\alpha_p-1}c\sigma_{pp}}{3\alpha_p(\alpha_p-1)D^2}\left(\frac{E_\gamma}{m_ec^2}\right)^{1-\alpha_p}
\]
\[
\times \int_0^{r_\tr{max}}\tr{d}r r^2n_N(r)\frac{\tilde{C}_p(r)}{(m_ec^2)^{\alpha_p-1}}
\]
Encompassing the $\gamma$-ray upper limit in the following shorthand
\begin{equation}
    \mathcal{F}=F_{\gamma,\tr{max}}(>E_\gamma)\left(\frac{E_\gamma}{m_ec^2}\right)^{\alpha_p-1},
\end{equation}
we can express this constraint as
\begin{equation}\label{eq:upperlimit}
\int_0^{r_\tr{max}}\tr{d}r r^2n_N(r)\frac{\tilde{C}_p(r)}{(m_ec^2)^{\alpha_p-1}}\leq 
\frac{3\alpha_p(\alpha_p-1)D^2\mathcal{F}}{2\xi K_\pi^{\alpha_p-1}c\sigma_{pp}} 
\end{equation}
This inequality holds for all upper limits on $F_\gamma$, so the smallest value of $F_{\gamma,\tr{max}}(>E_\gamma)(E_\gamma/m_ec^2)^{\alpha_p-1}$ across all observations gives us the tightest constraint. Since we have already predicted $\tilde{C}_p(r)$ in the primary- and secondary-dominated limits, we can just plug it into the above and interpret the results.

\subsubsection{Primary limit}
In section \ref{sec:primlim} we derived a relationship \eqref{eq:primlim} between the CRp distribution function $\tilde{C}_p(r)$ and the synchrotron power $S_\nu$ in the limit where CRe are dominated by primaries. By combining this with the above upper limit on $\tilde{C}_p(r)$ from $\gamma$-ray observations, again taking $\alpha_p=2s$, we obtain:
\[
\int_0^{r_\tr{max}}\tr{d}r r^2n_N(r) \frac{\zeta\sigma_TcB_0D^2}{r^2v_t(r)e^3}\frac{(2s-1)\mathcal{S}\mathcal{I}_2(r)}{6\pi\mathcal{N}\mathcal{I}_1}\leq
\]
\[
\frac{3s(2s-1)D^2\mathcal{F}}{\xi K_\pi^{2s-1}c\sigma_{pp}},
\]

\begin{equation}
\int_0^{r_\tr{max}}\tr{d}r\frac{n_N(r)}{v_t(r)}\mathcal{I}_2(r)\leq\frac{18\pi s\mathcal{N}\mathcal{I}_1\mathcal{F}e^3}{\zeta\xi K_\pi^{2s-1}B_0\sigma_T\sigma_{pp}c^2\mathcal{S}}
\end{equation}

Ultimately this amounts to a constraint on the transport speed $v_t$ for a given model and set of observations. In the very special case of a spatially-independent $v_t$ we get a simple lower limit
\begin{equation}\label{eq:advlim}
v_t\geq\frac{\zeta\xi K_\pi^{2s-1}B_0\sigma_T\sigma_{pp}c^2\mathcal{S}}{18\pi s\mathcal{N}\mathcal{I}_1\mathcal{F}e^3}\int_0^{r_\tr{max}}\tr{d}rn_N(r)\mathcal{I}_2(r)
\end{equation}
We evaluate this limit for different cluster models and discuss its characteristics in section \ref{sec:results}. It should be kept in mind that (\ref{eq:advlim}) is a radial transport speed and is perhaps most easily interpreted if the magnetic field is relatively coherently oriented. If the field is highly tangled, diffusion is a better model.

Suppose instead we consider diffusive transport characterized by a constant diffusion coefficient $\kappa$ as described by equation \eqref{eq:crpdiff}. Plugging this into the upper limit \eqref{eq:upperlimit} gives
\[
\int_0^{r_\tr{max}}\tr{d}r r^2n_N(r)\frac{\zeta\sigma_TcB_0D^2}{\kappa e^3}\frac{(2s-1)\mathcal{S}\mathcal{I}_3(r)}{6\pi\mathcal{N}\mathcal{I}_1}
\]
\[
\leq\frac{3s(2s-1)D^2\mathcal{F}}{\xi K_\pi^{2s-1}c\sigma_{pp}}
\]
which, finally, can be rearranged into a lower limit on the diffusion coefficient:
\begin{equation}\label{eq:kappalimit}
\kappa\geq\frac{\zeta\xi K_\pi^{2s-1}B_0\sigma_T\sigma_{pp}c^2\mathcal{S}}{18\pi s\mathcal{N}\mathcal{I}_1\mathcal{F}e^3}\int_0^{r_\tr{max}}\tr{d}rr^2n_N(r)\mathcal{I}_3(r)
\end{equation}
We evaluate this limit for different cluster models and discuss its characteristics in section \ref{sec:results}.

\subsubsection{Secondary limit}\label{sec:gammapred}
In the limit of secondary CRe dominating the population, we found that the radio observations directly translate into a CRp distribution \eqref{eq:seclim}. The $\gamma$-ray upper limit then offers a consistency check - if the inequality above is violated, secondaries cannot dominate the CRe population. Plugging \eqref{eq:seclim} into the upper limit \eqref{eq:upperlimit}, we obtain
\[
\frac{(2s-1)\mathcal{S}}{4\xi\pi\mathcal{N}\mathcal{I}_1}\left(\frac{4}{K_\pi}\right)^{2s-1}\frac{\sigma_T}{\sigma_{pp}}\frac{B_0D^2}{e^3}\int_0^{r_\tr{max}}\tr{d}rr^2\eta_e(r)\frac{B^2(r)+B_\tr{cmb}^2}{B_0^2}
\]
\[
\leq\frac{3s(2s-1)D^2\mathcal{F}}{\xi K_\pi^{2s-1}c\sigma_{pp}},
\]
\begin{equation}
\frac{\mathcal{I}_2(r_\tr{max})}{\mathcal{I}_1}\leq \frac{12\pi s\mathcal{N}}{4^{2s-1}}\frac{1}{c\sigma_T}\frac{e^3}{B_0}\frac{ \mathcal{F}}{\mathcal{S}}
\end{equation}

We can of course frame this in another way, using the secondary-dominated assumption to make a prediction for the $\gamma$-ray number flux:
\begin{equation}\label{eq:fluxpred1}
F_{\gamma}(>E_\gamma)=\frac{\mathcal{S}\mathcal{I}_2(r_\tr{max})}{12\pi s\mathcal{N}\mathcal{I}_1}\left(\frac{E_\gamma}{4m_ec^2}\right)^{1-2s}\frac{B_0}{e^3}c\sigma_T,
\end{equation}
or equivalently, a prediction for the $\gamma$-ray energy flux in some energy band
\begin{equation}\label{eq:fluxpred2}
\Phi_\gamma(E_1,E_2)=F_\gamma(>E_2)\frac{2s-1}{2s-2}\left(\left(\frac{E_2}{E_1}\right)^{2s-2}-1\right)
\end{equation}
to be directly compared with measurements such as from \cite{fermi16}. See section \ref{sec:results} for discussion on this limit. Note that since invoking the secondary limit for CRe tells us about the CRp density directly, the above prediction does not depend on any transport parameters, or on $\zeta$, $K_\pi$, or $\xi$.

\section{Results and Dependencies}\label{sec:results}
We consider here the results of our analysis in the various limiting cases in turn and discuss their dependence on different model parameters.

\subsection{Primary Limit, Advective Transport}\label{sec:primres}
In the limit of primary CRe dominating over secondaries, we found a lower limit on the advective transport speed given by equation \eqref{eq:advlim}. To evaluate this, we model the Coma cluster using the density model $n_e$ given in \S\ref{shape}. The total mass of this model is divergent, so we must also assume a cluster extent $r_\tr{max}$. Based on the X-ray data from \cite{briel92} we estimate a maximum radius of $r_\tr{max}=4000$ kpc. This choice is somewhat arbitrary, but as we will see the results do not vary much within the range of reasonable values of $r_\tr{max}$. As a fiducial magnetic field model we take $\alpha_B=0.5$ and $B_0=3\ \mu$G. We will anchor the radio observations at 1.4 GHz with measured intensity $S_{\tr{1.4 GHz}}\approx 6.4\ee{-24}$ erg cm$^{-2}$ s$^{-1}$ Hz$^{-1}$ (\cite{deiss97}), and take the spectral index to be $s=1.35$ (see \cite{thierbach03}).

For the $\gamma$-ray flux limits, we impose the energy flux limit in the 7.5 - 10 GeV band from \cite{fermi16} of $\approx 10^{-7}$ MeV cm$^{-2}$ s$^{-1}$ (the exact value of the limit depends on the spatial model of the emission). We use the highest energy band so as to be most in line with our assumption of high energy. For a power law spectrum of index $2s$, this translates into a photon number flux limit above 10 GeV of
\begin{equation}
F_{\gamma,\tr{max}}=\frac{(2s-2)10^{-11}}{(2s-1)(1.33^{2s-2}-1)} \tr{photons cm}^{-2}\tr{ s}^{-1}
\end{equation}
This will go into the $\mathcal{F}$ quantity in our analysis.

We assume a relative acceleration efficiency of $\zeta=100$, but emphasize that we merely use this as a reference value. This parameter is informed by observations of galactic CRs, but in principle is highly uncertain. The acceleration mechanisms in a galaxy cluster such as Coma may be 
highly varied, and may include galactic, AGN, and ICM sources. We have purposefully avoided any treatment of specific mechanisms and encompassed all our ignorance of them into $\zeta$\footnote{By picking a single value in this way we are essentially assuming the nature of the acceleration is the same at all locations in the cluster. This is hardly certain, but any treatment of acceleration mechanisms beyond this simplification is outside the scope of this work.}. Fortunately the dependence of our transport limits $v_t$ and $\kappa$ on this parameter are linear, and so any uncertainty in $\zeta$ can be propagated easily. This analysis can also be turned around to impose constraints on $\zeta$ (and consequently on the acceleration mechanisms present) if we fix all other model parameters. We discuss this possibility in \S\ref{sec:conc}.

For the above fiducial values, we derive a lower limit on bulk CRp transport of $v_t\geq 42$ km s$^{-1}$. The dependence of this limit on some of the model parameters is explicit from equation \eqref{eq:advlim}: $v_t$ is directly proportional to the proton-to-electron relative acceleration efficiency $\zeta$ and the total radio surface brightness $S_\nu$, and inversely proportional to the upper limit on $\gamma$-ray flux $F_{\gamma,\tr{max}}$. It depends on the normalization of the magnetic field as $B_0^{1-s}$ for strong fields ($B_0\gg B_\tr{cmb}$) and as $B_0^{-1-s}$ for weak fields. If we assume $\xi$ and $K_\pi$ are related by the observational constraint that the total fraction of CRp kinetic energy in each collision, $\xi K_\pi$, is constant, we see that $v_t\propto K_\pi^{2s-2}$.

\begin{figure}
    \centering
    \includegraphics[width=0.5\textwidth]{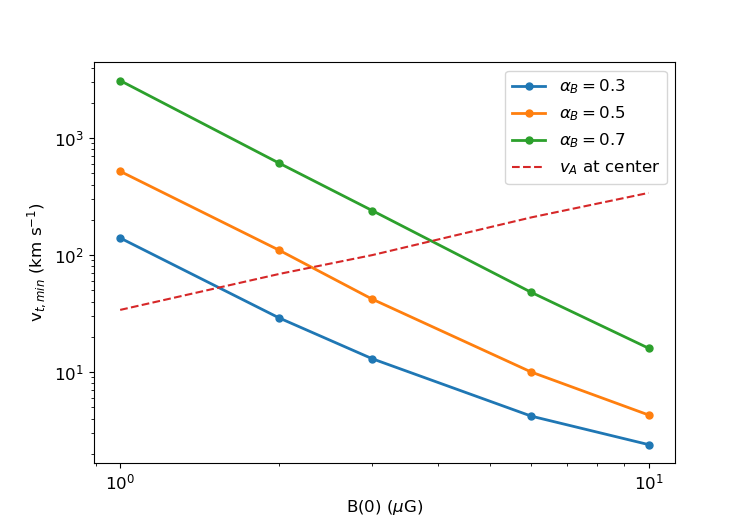}
    \caption{Minimum net outward flow velocity of CRp required to reduce pion decay $\gamma$-ray emission below current upper limits, assuming CRe secondaries are negligible. Results shown for different cluster magnetic field models. The central Alfv\'en speed for each model is shown for reference.}
    \label{fig:vt1}
\end{figure}

The dependencies on other parameters such as cluster extent $r_\tr{max}$, radio spectral index $s$, and magnetic field-density dependence $\alpha_B$ are not explicit, so we try many different models to find empirical trends. The minimum speed $v_t$ for various magnetic field models is shown in figure \ref{fig:vt1}. $v_t$ seems to be most sensitive to $B_0$, roughly following the relation
\begin{equation}
v_t\approx 42 \left(\frac{\zeta}{100}\right)\left(\frac{B_0}{3\ \mu\tr{G}}\right)^{-1-s}\tr{ km s}^{-1}.
\end{equation}

A complete table of the results is shown in Table \ref{tab:table}. This table shows the effects of changing the model parameters with non-explicit dependence, compared to the fiducial model. We see that changing $B_0$ from 1 to 10 decreases the resulting $v_\tr{t,min}$ by a factor of about 100 or more, while changing $\alpha_B$ from 0.3 to 0.7 results in an increase of $v_\tr{t,min}$ by about a factor of 10 (see first block of Table \ref{tab:table}. In contrast, changing the shape parameter $\alpha_\eta$ from 0 (uniform CRe density) to 1 (CRe density proportional to $n_e$) decreases $v_\tr{t,min}$ by less than a factor of 2 (see second block of Table \ref{tab:table}). This is far less than the dependence on $B$ or on the proton-to-electron relative acceleration efficiency $\zeta$, and suggests that our analysis is not sensitive to our choice of the form of the CRe shape function.

There is only slight dependence of $v_t$ with the other parameters in Table \ref{tab:table} except perhaps with the assumed extent of the cluster $r_\tr{max}$ (see fourth block of Table \ref{tab:table}). Physically this is because a larger cluster has greater gas content, so CRp densities would have to be lower to avoid a detectable $\gamma$-ray signal. But, as the size of the Coma cluster is almost certainly between 3 and 5 Mpc, the possible variation in $v_\tr{t,max}$ due to uncertainty in this parameters is relatively small.

\begin{table*}\label{tab:table}
\begin{tabular}{l l l l l l l l l | l l l}
\hline
$B_0^a$&$\alpha_B^b$&$\alpha_\eta^c$&$s^d$&$r_\tr{max}^e$&$r_c^f$&$\beta^g$&$K_\pi^h$&$\xi^i$&$v_\tr{t,min}^j$&$\kappa_\tr{min}^k$&$\Phi_\gamma^l$(7.5-10 GeV)\\
($\mu$G)&&&&(Mpc)&(kpc)&&&&(km s$^{-1}$)&(10$^{30}$ cm$^2$s$^{-1}$)&(10$^{-10}$ GeV cm$^{-2}$s$^{-1}$)\\
\hline 
1 & 0.3& 0.85& 1.35& 4& 294& 0.75& 0.17& 3& 140 & 65  &29  \\
2 & 0.3& 0.85& 1.35& 4& 294& 0.75& 0.17& 3& 29  & 13  &5.9 \\
3 & 0.3& 0.85& 1.35& 4& 294& 0.75& 0.17& 3& 13  & 5.7 &2.4 \\
6 & 0.3& 0.85& 1.35& 4& 294& 0.75& 0.17& 3& 4.2 & 1.6 &0.60\\
10& 0.3& 0.85& 1.35& 4& 294& 0.75& 0.17& 3& 2.4 & 0.84&0.27\\
1 & 0.5& 0.38& 1.35& 4& 294& 0.75& 0.17& 3& 520 & 350 &250 \\
2 & 0.5& 0.38& 1.35& 4& 294& 0.75& 0.17& 3& 110 & 70  &50  \\
3 & 0.5& 0.38& 1.35& 4& 294& 0.75& 0.17& 3& 42  & 27  &19  \\
6 & 0.5& 0.38& 1.35& 4& 294& 0.75& 0.17& 3& 10  & 6.0 &4.0 \\
10& 0.5& 0.38& 1.35& 4& 294& 0.75& 0.17& 3& 4.3 & 2.2 &1.3 \\
1 & 0.7&-0.09& 1.35& 4& 294& 0.75& 0.17& 3& 3100& 2600&2700\\
2 & 0.7&-0.09& 1.35& 4& 294& 0.75& 0.17& 3& 610 & 510 &530 \\
3 & 0.7&-0.09& 1.35& 4& 294& 0.75& 0.17& 3& 240 & 200 &210 \\
6 & 0.7&-0.09& 1.35& 4& 294& 0.75& 0.17& 3& 48  & 39  &41  \\
10& 0.7&-0.09& 1.35& 4& 294& 0.75& 0.17& 3& 16  & 12  &12  \\
\hline 
3 & 0.5& 0.0 & 1.35& 4& 294& 0.75& 0.17& 3& 51  & 41  &40  \\
3 & 0.5& 0.2 & 1.35& 4& 294& 0.75& 0.17& 3& 47  & 34  &28  \\
3 & 0.5& 0.38& 1.35& 4& 294& 0.75& 0.17& 3& 42  & 27  &19  \\
3 & 0.5& 0.6 & 1.35& 4& 294& 0.75& 0.17& 3& 37  & 20  &12  \\
3 & 0.5& 1.0 & 1.35& 4& 294& 0.75& 0.17& 3& 30  & 12  &4.3 \\
\hline 
3 & 0.5& 0.48& 1.15& 4& 294& 0.75& 0.17& 3& 80  & 48  &7.4 \\
3 & 0.5& 0.43& 1.25& 4& 294& 0.75& 0.17& 3& 58  & 36  &12  \\
3 & 0.5& 0.38& 1.35& 4& 294& 0.75& 0.17& 3& 42  & 27  &19  \\
3 & 0.5& 0.33& 1.45& 4& 294& 0.75& 0.17& 3& 31  & 21  &31  \\
3 & 0.5& 0.28& 1.55& 4& 294& 0.75& 0.17& 3& 23  & 16  &48  \\
\hline 
3 & 0.5& 0.38& 1.35& 1 & 294& 0.75& 0.17& 3& 14  & 1.8 &1.8\\
3 & 0.5& 0.38& 1.35& 2 & 294& 0.75& 0.17& 3& 24  & 7.1 &5.4\\
3 & 0.5& 0.38& 1.35& 4 & 294& 0.75& 0.17& 3& 42  & 27  &19 \\
3 & 0.5& 0.38& 1.35& 8 & 294& 0.75& 0.17& 3& 75  & 100 &76 \\
3 & 0.5& 0.38& 1.35& 16& 294& 0.75& 0.17& 3& 130 & 380 &310\\
\hline 
3 & 0.5& 0.38& 1.35& 4& 100& 0.75& 0.17& 3& 35  & 25  &160 \\
3 & 0.5& 0.38& 1.35& 4& 200& 0.75& 0.17& 3& 39  & 27  &41  \\
3 & 0.5& 0.38& 1.35& 4& 294& 0.75& 0.17& 3& 42  & 27  &19  \\
3 & 0.5& 0.38& 1.35& 4& 400& 0.75& 0.17& 3& 45  & 28  &11  \\
3 & 0.5& 0.38& 1.35& 4& 500& 0.75& 0.17& 3& 47  & 28  &7.2 \\
\hline 
3 & 0.5& 0.22& 1.35& 4& 294& 0.55& 0.17& 3& 55  & 31  &10  \\
3 & 0.5& 0.31& 1.35& 4& 294& 0.65& 0.17& 3& 51  & 31  &15  \\
3 & 0.5& 0.38& 1.35& 4& 294& 0.75& 0.17& 3& 42  & 27  &19  \\
3 & 0.5& 0.43& 1.35& 4& 294& 0.85& 0.17& 3& 33  & 22  &21  \\
3 & 0.5& 0.47& 1.35& 4& 294& 0.95& 0.17& 3& 25  & 16  &20  \\
\hline 
3 & 0.5& 0.38& 1.35& 4& 294& 0.75& 0.50& 1& 88  & 57  &19  \\
3 & 0.5& 0.38& 1.35& 4& 294& 0.75& 0.25& 2& 54  & 35  &19  \\
3 & 0.5& 0.38& 1.35& 4& 294& 0.75& 0.17& 3& 42  & 27  &19  \\
3 & 0.5& 0.38& 1.35& 4& 294& 0.75& 0.13& 4& 36  & 23  &19  \\
3 & 0.5& 0.38& 1.35& 4& 294& 0.75& 0.10& 5& 29  & 19  &19  \\
\hline
\end{tabular}
\caption{Results for varying different model parameters with non-explicit dependencies. $^a$Central magnetic field strength, defined in \eqref{eq:bfield} $^b$Magnetic field dependence on gas density, defined in \eqref{eq:bfield}. $^c$Shape function dependence on gas density, defined in \eqref{eq:shape1}. $^d$Spectral index of radio emission. $^e$Maximum extent of cluster. $^f$Cluster core radius, defined in \eqref{eq:beta}. $^g$Exponent of gas density profile, defined in \eqref{eq:beta}. $^h$Fraction of initial CRp kinetic energy contained in each pion produced in hadronic reactions, defined in \eqref{eq:pionE}. $^i$Pion multiplicity (number of pions produced per hadronic reaction), defined in \eqref{eq:pionE}. $^j$Minimum velocities according to \eqref{eq:advlim}. $^k$Minimum diffusion coefficients according to \eqref{eq:kappalimit}. $^l$Gamma ray energy flux predicted in the 7.5-10 GeV band, assuming CRe are dominated by secondaries, given by \eqref{eq:fluxpred2}.}
\end{table*}

\subsection{Primary Limit, Diffusive Transport}\label{subsec:primdiff}
\begin{figure}
\includegraphics[width=0.5\textwidth]{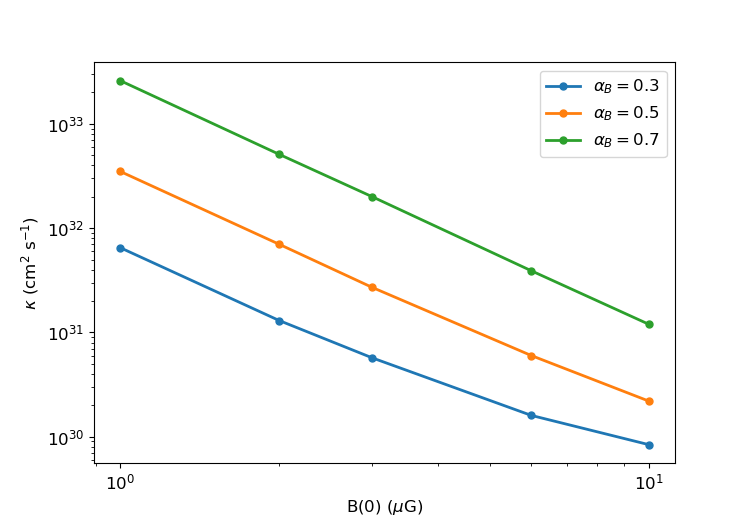}
\caption{Minimum radial diffusion coefficient required to reduce $\gamma$-ray emission below current upper limits, assuming CRe secondaries are negligible. Results shown for different cluster magnetic field models.}\label{fig:k1}
\end{figure}

If we describe CRp transport as a diffusive process with diffusion coefficient $\kappa$ that is constant over the (relatively small) observationally constrained range we find the $\gamma$-ray upper limits put a lower limit on $\kappa$ according to \eqref{eq:kappalimit}. For the fiducial values described above, this lower limit comes out to $2.7\ee{31}$ cm$^2$/s. We plot the values of $\kappa$ for different models in figure \ref{fig:k1}. Again the strongest dependence is on $B_0$, following the same scaling as $v_t$:
\begin{equation}
\kappa\approx 2.7\ee{31}\left(\frac{\zeta}{100}\right)\left(\frac{B_0}{3\ \mu\tr{G}}\right)^{-1-s}\tr{ cm}^2\tr{ s}^{-1}.
\end{equation}
The results of changing the other parameters are again shown in Table \ref{tab:table}. As with $v_\tr{t,min}$ the dependencies of $\kappa$ on parameters other than the magnetic field are significantly subdominant.

We can convert these limits on diffusion coefficient $\kappa$ into limits on the mean free path due to scattering seen by individual CRs $l_\tr{mfp}$ by the simple relation $l_\tr{mfp}=3\kappa/c$. For our fiducial model, $l_\tr{mfp}$ must be at least 900 pc for CRp to be able to leave the cluster fast enough to bring $\gamma$-radiation under the detection limit in Coma.

In a model where radial transport is due to scattering by magnetic inhomogeneities at the cosmic ray gyroscale, and the background magnetic fieldlines are nearly radial and straight, the advection speed $v_t$ and diffusion coefficient $\kappa$ can be combined to produce a lengthscale $R\equiv\kappa/v_t$ which is representative of the size of the system; for the values given here, $R\sim$ 2 Mpc, which is reasonable for Coma (the general expression for this quantity is
\[
\frac{\kappa}{v_t}=\frac{\int_0^{r_\tr{max}}\tr{d}rr^2n_N(r)\mathcal{I}_3(r)}{\int_0^{r_\tr{max}}\tr{d}rn_N(r)\mathcal{I}_2(r)},
\]
where $\mathcal{I}_2$ and $\mathcal{I}_3$ are defined in equations \eqref{eq:i2} and \eqref{eq:i3}).

\subsection{Magnetic Field Tangling}
Although our aim in this work is to put limits on the transport of CRs in a \emph{general} framework, it is useful to consider these limits in the context of specific transport models and see what they imply. For example, we could attribute the transport described above to turbulent motions of the gas in the Coma cluster. If we describe the turbulence by a length scale $L_\tr{MHD}$, where the characteristic speed is $v_A$, the eddy motions will induce diffusion of order
\[
\kappa_\tr{turb}\sim \frac{L_\tr{MHD}v_A}{3}.
\]
However, for the typical Alfv\'en speeds of $\sim 100$ km s$^{-1}$ this would require $L_\tr{MHD}$ to be as large as the Coma cluster itself. For turbulence this strong, wave damping effects would also be important, and it is likely that streaming motions would outpace any diffusion from turbulence (see \cite{wiener13a}). Put another way, the diffusivities estimated in Table \ref{tab:table} are much larger than the eddy diffusivity corresponding to large scale cluster turbulence unless the turbulent length scale is of order the global size of the cluster.

Alternatively, motivated by observations which show that cluster magnetic fields have strong random components (\cite{vogt05}, \cite{bonafede10}) we can interpret $\kappa$ as the result of the random walk of CRs along a magnetic field which is tangled on a lengthscale $l_\tr{corr}$ (this works if $l_\tr{corr}$ is intermediate between the size of the system and the cosmic ray gyroradius). In this case, the maximum CR diffusion rate is the one given by field line random walk at the speed of light (\cite{minnie09}): $\kappa\sim l_\tr{corr}c/3$. This is also approximately the maximum rate of transport across magnetic fieldlines found in \cite{desiati14}. Additional scattering up and down the tangled fieldlines, which could be represented by a streaming velocity such as $v_s<c$, only reduces the transport rate (\cite{rechester78}). Therefore, the minimum tangling length $l_\tr{corr}$ consistent with the diffuse $\gamma$-ray upper limits is the same as the mean free path $l_\tr{mfp}$ estimated in the preceding section.

\begin{figure}
\includegraphics[width=0.5\textwidth]{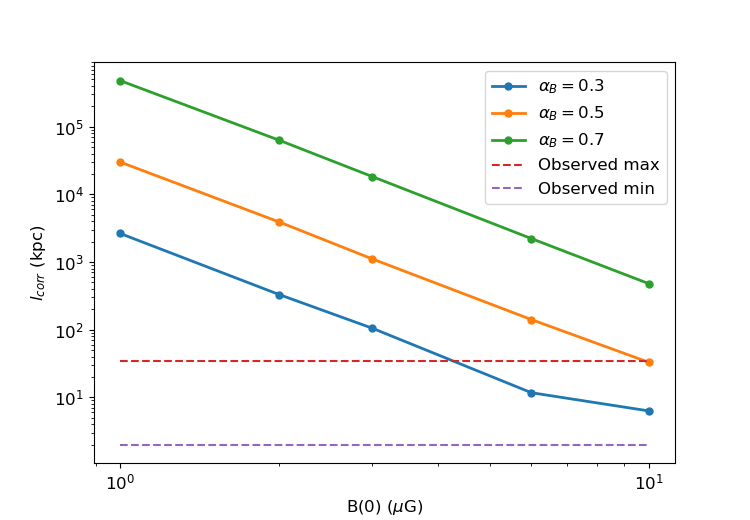}
\caption{Amount of magnetic field tangling allowed (in the self-confinement regime) while staying under $\gamma$-ray emission upper limits, assuming CRe secondaries are negligible and $\zeta=100$. If the field is tangled on scales less than this value, CRp cannot escape the cluster fast enough to be consistent with observations. Results shown for different magnetic field models. Reducing $\zeta$ to some value below 100 would shift these curves down by a factor of $\zeta/100$. The dotted lines represent the range of tangling lengths (2 - 34 kpc) in the magnetic turbulence in Coma (from \protect\cite{bonafede10}).}\label{fig:lcorr}
\end{figure}

To drive this point home, we have predicted what the streaming speed $v_s$ would be as a function of $r$ for the 100 GeV CRp that would be responsible for (undetected) $\gamma$-radiation for a specific self-confinement model. We calculate $v_s$ by taking the resulting CRp density profile from our model \eqref{eq:crpdiff} and plugging it into the formalism in \cite{wiener13a} and \cite{wiener18} for cosmic ray self-confinement with wave damping due to MHD turbulence, characterized by the length scale $L_\tr{MHD}=100$ kpc. According to this model, the bulk streaming speed $v_s$ of CRs along magnetic field lines is set by equating the growth rate of the CR streaming instability to the rate of damping by the background medium. This gives us $v_s$ as a function of position in the cluster, which can be translated into a minimum allowed field tangling length $l_\tr{corr}=3\kappa/v_s$, where $\kappa$ is the minimum diffusion coefficient derived from the observations according to our analysis in this work. We record the maximum (most restrictive) $l_\tr{corr}$ over all values of $r$ for each model and plot the results in figure \ref{fig:lcorr}, which also includes the range of length scales in the magnetic turbulence power spectrum inferred from Faraday rotation observations of Coma (\cite{bonafede10}).

We see from figure \ref{fig:lcorr} that the magnetic field observed in Coma is tangled too much ($l_\tr{corr}$ is too small) to explain the lack of $\gamma$-radiation under this set of assumptions (primary CRe dominate, CRp are transported via streaming with the assumed level of wave damping). This tension is relieved if the cluster magnetic field is in the upper range of values tested, which reduces the CRe population required to explain the radio emission, or if the relative proton-to-electron acceleration efficiency factor $\zeta$ is smaller than Milky Way values.

\subsection{Secondary Limit}
\begin{figure}
\includegraphics[width=0.5\textwidth]{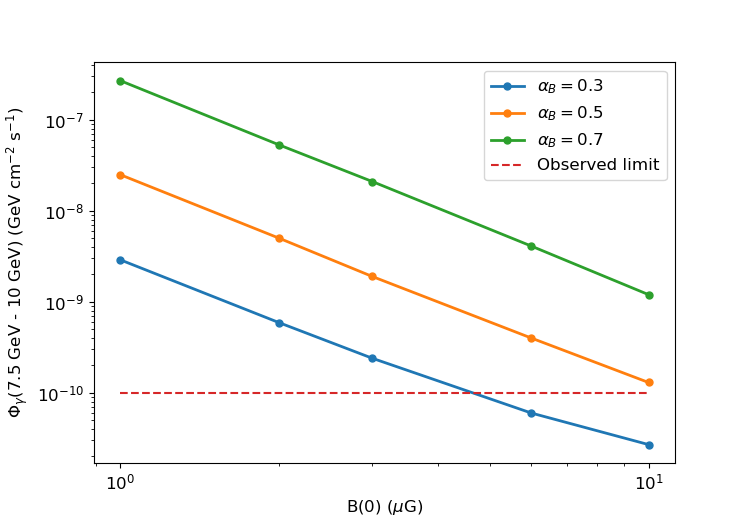}
\caption{Predicted $\gamma$-ray flux from pion decay in the 7.5-10 GeV band, assuming secondary CRe dominate. Results shown for different cluster magnetic field models. The dotted line represents the non-detection upper limit from \protect\cite{fermi16}. This model is only consistent with the $\gamma$-ray observations at the highest magnetic fields and flattest magnetic profiles.}\label{fig:flu}
\end{figure}

In the limit where the CRe are dominated by secondaries from hadronic interactions, we derived a flux prediction \eqref{eq:fluxpred2} which can be used to predict $\gamma$-ray energy fluxes in arbitrary energy bands. We compare with the highest energy band used in \cite{fermi16}, which is about 7.5 - 10 GeV. \cite{fermi16} find an upper limit in this bin of (depending on their emission model) around 6-10$\ee{-11}$ GeV cm$^{-2}$ s$^{-1}$. We compare this to the prediction of our models $\Phi_\gamma$ in figure \ref{fig:flu}. Variations of $\Phi_\gamma$ with all the non-explicit parameters are shown in the last column of Table \ref{tab:table}. Our fiducial model predicts a $\gamma$-ray flux of $1.9\ee{-9}$ GeV cm$^{-2}$ s$^{-1}$, far above the \emph{Fermi} upper limit. A secondary-only model for Coma's radio halo is thus highly disfavored.

\section{Consistency Checks}\label{sec:checks}
In this section we check some of the assumptions we have made in various regimes.
\subsection{CRe Transport Losses}
Throughout this work we have assumed that the CRe losses in the relevant energy range were dominated by synchrotron and IC emission. Namely, we neglected the transport term the CRe evolution equation (second term in equation \eqref{eq:transporteq}). We estimate these transport effects here and compare them to the radiative losses.

Suppose we can characterize the CRe transport with a bulk transport speed $v_t(r,\gamma_e)$, as we did for the CRp. Then the number loss (in particles per volume per unit $\gamma_e$ per time) due to transport is:
\[
\dot{f}_\tr{cre,trans}(r,\gamma_e)=\nabla\cdot(f_e(r,\gamma_e)v_t(r,\gamma_e))
\]
\[
=\gamma_e^{-\alpha_e}\frac{1}{r^2}\pp{}{r}(r^2C_e(r)v_t(r,\gamma_e))
\]

This is to be compared to the radiative loss rate (the third term in equation \eqref{eq:transporteq}),
\[
\dot{f}_\tr{cre,rad}(r,\gamma_e)=\pp{}{\gamma_e}(\dot{\gamma}_\tr{loss}f_e(r,\gamma_e))=s_e(r,\gamma_e)
\]
\[
=\gamma_e^{1-\alpha_e}G_e(r)=\gamma_e^{1-\alpha_e}\frac{\alpha_e-2}{6\pi}\frac{\sigma_Tc}{m_ec^2}C_e(r)(B^2(r)+B_\tr{cmb}^2)
\]
In the above we have assumed a power law for the CRe distribution, $\dot{\gamma}_\tr{loss}$ refers to the radiative losses per electron \eqref{eq:radloss}, and $G_e(r)$ refers specifically to the solution from our previous analysis (\ref{eq:esource0},\ref{eq:esource}).

In the simple case of a constant, energy-independent transport speed $v_t(r,\gamma_e)=v_t$, the ratio of transport losses to radiative losses for CRe is
\[
\frac{\dot{f}_\tr{cre,trans}(r,\gamma_e)}{\dot{f}_\tr{cre,rad}(r,\gamma_e)}=\frac{1}{\gamma_e}\frac{6\pi}{\alpha_e-2}\frac{v_t}{c}\frac{m_ec^2}{B^2(r)+B^2_\tr{cmb}}\frac{1}{\sigma_T}\left[\frac{2}{r}+\frac{C'_e(r)}{C_e(r)}
\right]
\]
Using our estimate of the CRe shape function $\eta_e(r)=n_e^{\alpha_\eta}(r)=(1+(r/r_c)^2)^{-3\alpha_\eta\beta/2}$, we get
\[
\frac{C'_e(r)}{C_e(r)}=\frac{\eta'_e(r)}{\eta_e(r)}=-\frac{3}{2}\beta\alpha_\eta\frac{2r}{r_c^2}\frac{1}{1+(r/r_c)^2}=-\frac{2}{r}\frac{3\beta\alpha_\eta r^2}{2(r^2+r_c^2)}
\]
and so the transport to radiative loss ratio is
\[
\frac{\dot{f}_\tr{cre,trans}(r,\gamma_e)}{\dot{f}_\tr{cre,rad}(r,\gamma_e)}=\frac{1}{\gamma_e}\frac{6\pi}{\alpha_e-2}\frac{v_t}{c}\frac{m_ec^2}{B^2(r)+B^2_\tr{cmb}}\]
\[
\times\frac{2}{\sigma_Tr}\left[1-\frac{3\beta\alpha_\eta r^2}{2(r^2+r_c^2)}\right]
\]
\[
=\frac{1}{\gamma_e}\frac{6\pi}{\alpha_e-2}\frac{v_t}{c}\frac{m_ec^2}{B^2(r)+B^2_\tr{cmb}}\frac{2}{\sigma_Tr}\frac{(2-3\beta\alpha_\eta)r^2+2r_c^2}{2(r^2+r_c^2)}
\]
The last term is a fairly weak function of $r$ of order unity, so let us discard it. Picking some scale values, the rest of the expression comes to
\begin{equation}
\frac{\dot{f}_\tr{cre,trans}(r,\gamma_e)}{\dot{f}_\tr{cre,rad}(r,\gamma_e)}\approx 370\frac{v_t}{c}\frac{10^4}{\gamma_e}\frac{r_c}{r}\frac{(4.4\ \mu\tr{G})^2}{B^2(r)+B^2_\tr{cmb}}.
\end{equation}
Recall that $r_c=297$ kpc for the Coma cluster. The scale field 4.4 $\mu$G is chosen as the effective combined field of a 3 $\mu$G field with $B_\tr{cmb}=3.24\ \mu$G.

We see that our assumption that this quantity is much much less than one is valid except when the transport speed $v_t$ is very large, or $r$ is very small. The approximation is better for higher energies.

\subsection{CRp Adiabatic Losses}
As mentioned in \S\ref{subsec:general}, energy losses to processes like Coulomb scattering and hadronic collisions are small for CRp above 1 GeV. But adiabatic losses may be significant. It is possible for CRs to be transported without suffering adiabatic losses, such as in a free-streaming model where the magnetic field is completely static. But if CR transport is governed by wave scattering from self-generated turbulence, as in \S\ref{subsec:specific}, we would expect significant adiabatic losses to be present which have not been accounted for in our analysis.

A detailed treatment describing how adiabatic loss times compare with transport times in the most general sense is beyond the scope of this paper. Instead we consider our limiting cases of advection with constant outward velocity $v_t$ and diffusion with constant coefficient $\kappa$. If the transport is diffusion-dominated, then there are no associated adiabatic losses and our analysis is correct as is.

Consider instead the advective limit for the specific case of constant $v_t$. If we suppose adiabatic losses are of the same order as expected from the specific model in \S\ref{subsec:specific}, then we have, using equation \eqref{eq:adiabatic} with $\mb{u}=v_t\hat{r}$,
\[
\dot{E}_\tr{adia}\sim \frac{E}{3}\nabla\cdot\mb{u}\sim \frac{2v_t}{3r}E.
\]
The adiabatic loss time $t_\tr{adia}=E/\dot{E}_\tr{adia}$ is then
\[
t_\tr{adia}=\frac{3r}{2v_t}.
\]
While loss times are larger than transport times $r/v_t$, they are of the same order. Our derivation of equation \eqref{eq:steadypadv}, which inherently included the assumption that adiabatic losses were negligible, would not be valid in this case (it would still be valid for a transport model with no adiabatic losses).

While we do not calculate the quantitative effect adiabatic losses on our results, the qualitative result of including these losses is clear. The total loss rate of energy in CRp is dictated by the observations. In our analysis we attribute all of this loss to transport. If some energy is lost through other processes, the transport need not be as fast as found in \S\ref{sec:results} to be consistent with observations. The limits derived in \S\ref{sec:results} would be relaxed, i.e. the values of $v_{t,\tr{min}}$ would be lower by some amount.

\subsection{Estimate of CRe Secondaries}
We have done our analysis in two limiting cases, where the CRe are dominated by primaries and secondaries respectively. We have already seen in \S\ref{sec:gammapred} that the upper limits on $\gamma$-ray flux rule out the secondary-dominated limit in Coma unless the magnetic field is fairly high. However, we have yet to check how many secondaries would be produced in the primary-dominated limit. If too many are produced, the primary-dominated limit assumption is inconsistent.

The CRp distribution in the primary-dominated limit inferred from radio observations was determined to be \eqref{eq:primlim}. The secondary CRe source function expected from a general CRp distribution was determined to be \eqref{eq:secsrc}. The expected secondary CRe source function in the primary-dominated limit is then
\[
s_{e,\tr{sec}}(r,\gamma_e)=\frac{8}{3}\frac{\xi}{K_\pi}\left(\frac{K_\pi}{4}\right)^{\alpha_p}c\sigma_{pp}\gamma_e^{-\alpha_p}n_N(r)\frac{\tilde{C}_p(r)}{(m_ec^2)^{\alpha_p-1}}
\]
\[
=\frac{8}{3}\frac{\xi}{K_\pi}\left(\frac{K_\pi}{4}\right)^{2s}c\sigma_{pp}\gamma_e^{-2s}n_N(r)\frac{\zeta\sigma_TcB_0D^2}{r^2v_t(r)e^3}\frac{(2s-1)\mathcal{S}\mathcal{I}_2(r)}{6\pi\mathcal{N}\mathcal{I}_1}
\]

This must be compared with the total CRe source function given by \eqref{eq:esource}. Their ratio is
\begin{multline}
\frac{s_{e,\tr{sec}}(r,\gamma_e)}{s_{e}(r,\gamma_e)}=\frac{8}{3}\frac{\xi}{K_\pi}\left(\frac{K_\pi}{4}\right)^{2s}\zeta\sigma_{pp}n_N(r)\frac{c}{v_t(r)}
\\
\times \left\lbrace\frac{\mathcal{I}_2(r)}{r^2\eta_e(r)\left(\frac{B^2(r)+B^2_\tr{cmb}}{B_0^2}\right)}\right\rbrace
\end{multline}
This can be found numerically, but we can get a good estimate by recognizing that the quantity in the braces is of order $r$:
\[
\frac{s_{e,\tr{sec}}(r,\gamma_e)}{s_{e}(r,\gamma_e)}\sim=\frac{8}{3}\frac{\xi}{K_\pi}\left(\frac{K_\pi}{4}\right)^{2s}\zeta\sigma_{pp}\mu_en_0\frac{c}{v_t(r)}r\left(1+\frac{r^2}{r_c^2}\right)^{-3\beta/2}
\]
For $s$ close to 1, and a fully ionized medium, $(8/3)(\xi/K_\pi)(K_\pi/4)^{2s}\zeta\mu_e$ is approximately 10, and $\sigma_{pp}$ is approximately 50 mbarn. Picking scale values for $n_0$ and $r$, we have
\begin{equation}
\frac{s_{e,\tr{sec}}(r,\gamma_e)}{s_{e}(r,\gamma_e)}\sim 10^{-4}\frac{n_0}{10^{-3}\ \tr{cm}^{-3}}\frac{r}{100\ \tr{kpc}}\frac{c}{v_t(r)}\left(1+\frac{r^2}{r_c^2}\right)^{-3\beta/2}
\end{equation}

Considering how large we expect $c/v_t$ to be, this suggests that our assumption of primary CRe domination is \emph{inconsistent}. Even at the large end of the range of values of $v_t$ we found, around 100 km s$^{-1}$, we may expect some 50\% of CRe to be secondaries. For expected relative acceleration efficiencies and CR bulk transport speeds, secondary CRe injected from hadronic processes \emph{must} make up a significant fraction of the total CRe source.

What does this mean for our earlier analysis? Qualitatively, the presence of secondary CRe suggests that for the same level of radio emission, fewer primary CRe are necessary. That is, we can explain the radio emission with some primary CRe source function that is some fraction of the one we derived in \S\ref{sec:eloss}. This makes it easier to get under the $\gamma$-ray upper limits, implying we can get away with lower transport speeds $v_t$. A more quantitative treatment is beyond the scope of this work.

We can alternatively derive a new, more restrictive lower limit on the transport speed $v_t$ such that secondaries are negligible. Let's say we required that secondaries make up less than one percent of all CRe. Then the transport speed must be at least
\begin{equation}
v_t(r)\gsim 10^{-2}c\frac{n_0}{10^{-3}\ \tr{cm}^{-3}}\frac{r}{100\ \tr{kpc}}\left(1+\frac{r^2}{r_c^2}\right)^{-3\beta/2}.
\end{equation}
Such speeds are not expected and we can reasonably surmise that secondary CRe in the Coma cluster are non-negligible. We touch on the implications of this in the next section.

\section{Discussion and Conclusions}\label{sec:conc}
Diffuse non-thermal radio continuum emission detected in galaxy clusters shows that cosmic ray electrons are present, but, contrary to expectations, there is no evidence that cosmic ray protons are present. The purpose of this paper was to explore the implications of these observations and find limits for cosmic ray acceleration and propagation in galaxy clusters.

We presented an analysis of CR populations based on radio and $\gamma$-ray observations. Given a cluster thermal plasma density and magnetic field strength model, radio observations constrain the CRe population and, assuming steady state, the CRe source strength. $\gamma$-ray upper limits put upper limits on the CRp density present in the cluster. We then combined these results in two limiting cases. In the case where we assume secondary CRe production is negligible, we arrived at a minimum transport speed necessary to reduce the CRp density below the upper limits. In the opposite case where we assume primary CRe production is negligible, we obtain a parameter-dependent $\gamma$-ray flux prediction to be compared to the upper limits. These equations are general in that they can be applied to any cluster for which the assumptions of the model are reasonably well satisfied.

Using radio (\cite{thierbach03}) and $\gamma$-ray (\cite{fermi16}) observations of the Coma cluster we have found that under the assumption that all synchrotron-emitting CRe are directly accelerated, we require CRp to be transported outward at a bulk speed of 10-100 km s$^{-1}$ or with diffusion coefficient 10$^{31}$-10$^{32}$ cm$^2$ s$^{-1}$ in order to explain the lack of $\gamma$-ray detection for our fiducial set of cluster parameters. These speeds are sub-Alfv\'enic, so a self-confinement model with a magnetic field which is coherent on large scales is consistent with the radio and $\gamma$-ray observations. If the field is tangled on the scales of a few kpc, as suggested by \cite{bonafede10}, then CRs which free stream at the speed of light will experience diffusive transport with diffusion coefficient comparable to the values found here. So a free streaming model with a tangled field is also consistent with the observations for our choice of parameters.

Conversely, a self-confinement model with a tangled field is \emph{not} consistent with observations for our chosen parameters, as CRp would not be transported outward fast enough. To bring this model in line with observations requires changing our assumed parameters. In section \ref{sec:results} we presented the dependence of our results on these parameters, shown in detail in Table \ref{tab:table}. We can use observations of the magnetic field to constrain other parameters in our model, namely the relative acceleration efficiency $\zeta$. As mentioned, some studies suggest that the confinement time of CRs in galaxy clusters is of order a Hubble time. This is not incompatible with our steady state analysis here, as it does not imply that no CRs ever leave the cluster. However, with reduced outward transport we require CRp to be accelerated less efficiently than in our fiducial model in order to be compatible with both the radio and $\gamma$-ray observations. We find that our model Coma cluster, with field tangled at kpc scales (\cite{bonafede10}) and CR transport determined by self-confinement, is consistent with observations if the relative proton to electron acceleration efficiency is less than 15. The exact value of this limit depends on the level of assumed wave damping. Various regions of the (extensive) parameter space can be ruled out in this way.

Additionally, at the comparatively low transport speeds derived here we expect CRe secondaries to make a significant contribution to the population. We also found that unless the magnetic field in the center of the Coma cluster is around 10 $\mu$G or above, much higher than observed, the CRe population cannot be dominated by secondaries or else the $\gamma$-ray flux prediction exceeds upper limits. This suggests neither component, primaries or secondaries, can be consistently ignored. The observations of Coma can only be explained by a hybrid population of primary CRe which are being continuously injected into the cluster and secondary CRe which arise from hadronic interactions. A more detailed treatment is therefore necessary to examine CR transport in the Coma cluster. We leave such a treatment for future work, but speculate that the inclusion of a small secondary component in the primary-dominated analysis would alleviate the limits derived here: the presence of secondaries implies less acceleration of primary CRe is needed to produce the observed radio emission, which in turn implies CRp are produced at a slower rate.

Finally, we reiterate that while we have specified our results to the example of the Coma cluster, the analysis is general and can be used on any individual system reasonably approximated by spherical symmetry.

\noindent{Acknowledgements:}
We are happy to acknowledge discussions with Gianfranco Brunetti. Comments by anonymous referees helped improve the content and clarity of the presentation. JW and EGZ acknowledge support by NSF Grant AST-1616037, the WARF Foundation, and the Vilas Trust.

\bibliography{master_references2}

\end{document}